# Partition function and magnetization of two-dimensional Ising models in non-zero magnetic field: A semi-empirical approach

M V Vismaya[1] and M V Sangaranarayanan[*]

Department of Chemistry

Indian Institute of Technology -Madras Chennai 600036 India

Email: vismayamv1112@gmail.com

Email: sangara@iitm.ac.in

Abstract

The partition functions of ferromagnetic Ising models of square lattices in a finite magnetic field is deduced using topological considerations within a heuristic graph-theoretical approach. These equations are derived separately for low and high temperature regimes while Onsager's exact solution is obtained therefrom when the magnetic field is zero. The derived partition function equations here are almost similar to those given by Onsager, thus indicating a straight-forward protocol, even when the magnetic field is present. The spontaneous magnetization derived here using the Helmholtz free energy is identical with that arising from Onsager's exact solution. The partition functions lead to the known series expansions of the magnetization and zero-field susceptibility.

## 1. Introduction

The formulation of the partition function of two-dimensional Ising models in the presence of a magnetic field has been a challenging task, ever since the analysis of zero-field case due to the pioneering analysis of Peierls [1], Kramers and Wannier [2] and Onsager [3]. While the one-dimensional nearest neighbour Ising model is of pedagogical interest on account of the absence of phase transitions, the corresponding two-dimensional analogue has a rich critical behaviour with universal critical exponents for spontaneous magnetization, zero-field susceptibility and specific heat, the latter displaying a logarithmic singularity at the critical temperature. Although the exact location of the critical temperature for two-dimensional lattices is due to Kramers and Wannier [2], a complete thermodynamic analysis of the system by deriving the Helmholtz free energy, internal energy, absolute entropy, specific heat etc is due to Onsager [3]. The rigorous and mathematically complex formalism of Onsager [3] has subsequently been partially demystified by Kac and Ward [4], Kaufman and Onsager [5], Feynman [6], Kasteleyn [7], Glasser [8], Potts and Ward[9], Newell and Montroll [10] ,Vvodichenko [11] , Hurst and Green [ 12 ]and Berezin[13].

The renormalisation group [14] methodology has provided new insights into the magnitude of the critical temperature. It must be emphasized that although the closed-form expression for the partition function in non-zero fields has been elusive, important insights have emerged by invoking scaling and universality hypothesis of Widom [15] as well as powerful numerical analysis of series expansions, (cf. Domb and Green [16] and Guttmann [17] . However, finite lattice sizes ($N$) in this context are of little value since the occurrence of critical phenomena can only be obtained in the thermodynamic limit ( $N \rightarrow \infty$). As pointed out by Cipra [18], the Ising model problem is 'apparently very difficult, yet certainly extremely important'. While the general expressions for the field-dependent partition function and magnetization are not known *per se*, these do exist for a particular value of the magnetic field viz $H = \frac{i\pi}{2kT}$ [19].In a complementary manner, the exact partition function of three-dimensional Ising model at zero magnetic field too has been elusive as emphasized by Viswanathan et al [ 20] and Fisher and Perk [ 21] ; consequently, extrapolation of two-dimensional analogues in a straight-forward manner is beset with difficulties.

Another effective strategy to deduce the partition function for large two-dimensional lattices consists in the systematic enumeration of 'black-white' edges denoted as $A(p, q)$ and incorporating these in the corresponding partition functions, when the field is non-zero [22]. $A(p, q)$ denotes the number of colourings for which there are exactly p black sites and q black-white edges in a lattice of $N$ sites. This methodology has been carried out for deriving the partition functions for $N$ =16, 25 and 36 sites and if extension to infinite sites is accomplished, this is a promising approach [23,24] . Using a small angle approximation, Guuderault [ 25] has provided new analytical results for $H \neq 0$ and the quantitative agreement with simulation results has been clearly demonstrated. As mentioned by Goldenfeld, [ 26] "much effort has been expended to solve it in 3 dimensions in zero magnetic field, although a solution of the $d$ = 2, $H \neq 0$ case might be more significant."

It will be aesthetically pleasing to search for the field-dependent partition functions that seamlessly yield Onsager's exact solutions so that the already-available critical behaviour when the magnetic field is zero will remain valid *ipso facto.* With this objective in mind, we propose a semi-empirical approach using topological considerations in order to deduce the partition function for infinite sites. We demonstrate that the proposed methodology yields (i) Onsager's exact solution when the field is zero; (ii) series expansions for spontaneous magnetization;(iii) correct dependence of magnetization on the magnetic field and (iv) series expansions for zero-field susceptibility. In contrast to all earlier approaches, we deduce the spontaneous magnetization from the first derivative of the Helmholtz free energy with respect to the magnetic field.

2.Methodology

For a square lattice of $N = n^2$ spins consisting n rows and n columns, the total system Hamiltonian ( $H_T$) is represented as [28]

$$H_T = -J \sum_{<k,k'>} \sigma_k \, \sigma_{k'} - H \sum_k \sigma_k \tag{1}$$

where $\sigma_k = \pm 1$ are the Ising spin variables and $\langle k, k' \rangle$ runs over nearest neighbour pairs.

The first sum is over all the nearest neighbour pairs and the second sum is over all the lattice sites. $J$ denotes the nearest neighbour interaction energy and $H$ denotes the external magnetic field. In the parlance of graph theory, the sites are vertices while the nearest neighbour bonds are edges.

The partition function can be represented as

$$Q = \prod \mathrm{e}^{-K\sigma_k\sigma_{k'} - h\sigma_k} \tag{2}$$

where the dimensionless quantities $K = \frac{J}{kT}$ and $h = \frac{H}{kT}$.

The coupling term '$K$' involves nearest neighbour pair and the field term pertains to the total number of sites. Here we first consider the case $h \geq 0$.

In order to derive the field-dependent partition function, it is imperative to recall the following well-known equation arising from Onsager's exact solution for $H = 0$ (cf. [29])

$$\frac{1}{N} logQ = log\left[\sqrt{2}\, cosh(2K)\right] + \frac{1}{\pi} \int_0^{\frac{\pi}{2}} d\phi \, log\left[1 + \sqrt{1 - \kappa^2 \sin^2 \phi}\right] \tag{3}$$

with

$$\kappa = \frac{2\, sinh(2K)}{cosh^2(2K)} = \frac{2(1 - M_0^8)^{\frac{1}{4}}}{1 + (1 - M_0^8)^{\frac{1}{2}}} \tag{4}$$

$\kappa$ has a maximum value of unity that reflects the magnitude of the non-dimensional critical temperature $\left(\frac{J}{kT_c}\right)$.

An analogous equation connecting $\kappa\ (H \neq 0)$ and $M\ (H \neq 0)$ should also exist, if the partition function in the presence of magnetic fields is rendered available.

The spontaneous magnetization $M_0$ corresponding to Onsager's exact solution is

$$M_0 = \sqrt[8]{1 - sinh^{-4}(2K)} \tag{5}$$

For extending equation (3) to derive the partition function at $H \neq 0$, several options are available such as (i) modification of the arguments of the hyperbolic functions; (ii) altering the integrand and (iii) re-formulation of $\kappa$ as a function of $H$ and $K$. Nevertheless, all these options are not easily amenable in view of the partially available disjointed information e.g. low temperature series for magnetization $M\ (H, K)$ and high-temperature series for susceptibility $\chi\ (H, K)$.

We have employed the combinatorial approach advocated by Glasser [8] to deduce the partition function at two different regimes: $T \leq T_c$ and $T > T_c$. With the help of the usual hyperbolic function identities and the allowed value of the spin variables as $\pm 1$, it follows that $\sigma_k\sigma_{k'} = \eta$ and $\eta^2 = 1$. Further,

$$e^{-K\eta} = cosh(K) - \eta\, sinh(K) = (1 - tanh^2 K)^{-\frac{1}{2}}(1 - \eta\, tanh(K))$$

Employing the periodic boundary conditions pertaining to two-dimensions, the term containing the magnetic field becomes

$$e^{-\frac{h}{4}(\sigma_k+\sigma_{k'})} = cosh\left(\frac{h}{4}\right) - (\sigma_k + \sigma_{k'})\, sinh\left(\frac{h}{4}\right) = (1 - tanh^2 h)^{-\frac{1}{8}}[(1 - \sigma_k\, tanh\, h)(1 - \sigma_{k'}\, tanh\, h)]^{\frac{1}{8}}.$$

For brevity, we denote $\frac{H}{kT}$ as $h$ and define $x = tanh(K)$ and $y = tanh(h)$. Hence

$$Q_{(h\geq 0)} = (1 - x^2)^{-N}\, (1 - y^2)^{-\frac{N}{4}} \sum \prod (1 - x\, \sigma_k\, \sigma_{k'})\, [(1 - y\, \sigma_k)(1 - y\sigma_{k'})]^{\frac{1}{8}} \tag{6}$$

As mentioned earlier, the bilinear term denotes the nearest neighbour interaction energy while the magnetic field term includes the single spin value. The definition of $x$ and $y$ in terms of *tan hyperbolic* functions is preferable since (i) limiting cases are easier to verify and (ii) isomorphism with random cluster models [30] can be made use of.

It is of interest to recall the complexity arising due to the presence of the '$y$' term that denotes the dimensionless magnetic field. While $x$ represents the weight per edge (bond), $y$ indicates weight per site (vertex). The coupling between $x$ and $y$ as depicted by equation (6) alters the scenario completely and mere enumeration of closed graphs as in the case of zero magnetic fields is no longer adequate. The factor $[(1 - y\, \sigma_k)(1 - y\sigma_{k'})]^{\frac{1}{8}}$ incorporates the effect of the magnetic field on the spins. Further, this term is responsible for the breaking-up of the spin-flip symmetry- which exists if only pairwise interactions $(1 - x\, \sigma_k \sigma_{k'})$ were present. At $y$ =0, equation (6) is identical with equation (4) of [8].

$$Q_{(h\geq 0)} = 2^N (1 - x^2)^{-N}\, (1 - y^2)^{-\frac{N}{4}} \sum_{subgraphs} G(x, y) \tag{7}$$

$G(x, y)$ denotes all closed graphs and includes graphs with odd degree vertices (open paths) contributing terms with $x$ as well as $y$.The rationale behind this is the appearance of a new effective coupling constant between spins, which now depends on parameters $x$ (coupling) and $y$ (weight from single-site terms). It is well known that closed paths correspond to sequence of edges on the lattice thereby leading to the formation of a loop. On the other hand, open paths appear in presence of an external field. Hence when $y \neq 0$, both closed and appropriate open paths need to be enumerated.

Each subgraph (open as well as closed) is associated with a modified statistical weight that depend on both $x = tanh(K)$ (bond variable) and $y = tanh(h)$(site variable). Furthermore, every occurrence of an open-end point (vertex of odd degree) in the subgraph includes a factor of y.

To elucidate this subtle feature, we expand the terms in equation (6)

$$\prod_{\langle k,k'\rangle} (1 - x\, \sigma_k\, \sigma_{k'}) \approx 1 - x \sum_{\langle k,k'\rangle} \sigma_k \sigma_{k'} + x^2 \sum_{\langle k,k'\rangle < \langle k'',k'''\rangle} \sigma_k \sigma_{k'} \sigma_{k''} \sigma_{k'''} + ..$$

$$\prod_{\langle k,k'\rangle} (1 - y\, \sigma_k)(1 - y\sigma_{k'}) = \prod_{\langle k,k'\rangle} (1 - y\, \sigma_k - y\sigma_{k'} + y^2 \sigma_k \sigma_{k'})$$

$$\prod_{\langle k,k'\rangle}[(1-y\,\sigma_k)(1-y\sigma_{k'})]^{\frac{1}{8}} = = \prod_{\langle k,k'\rangle}(1-y\,\sigma_k - y\sigma_{k'} + y^2\sigma_k\sigma_{k'})^{\frac{1}{8}}$$

The linear term in $y$ is associated with single spins ($\sigma_k$ *and* $\sigma_{k'}$ )and depicts the manner in which the magnetic field impacts the local spin alignment. The $y^2\sigma_k\sigma_{k'}$ terms represent a coupling term involving all the nearest neighbour pairs $(k,k')$ and generates closed path segments. Upon including the site terms $(y)$ or symmetrically distributing their effect over edges, these in turn influence not only the local spin behaviour but also alters the interactions between spin pairs. Thus, the net interaction between the pair of spins is no longer dictated by $x$ alone, but by a new effective coupling function which we denote as $\boldsymbol{\alpha}(K,h)$ . This term incorporates the original nearest neighbour interaction as well as the effects generated by the presence of the magnetic field. This is because while expanding the products of edge and site factors, the initial $x\,\sigma_k\,\sigma_{k'}$ pair interactions and additional terms involving $y$ along with $\sigma_k\sigma_{k'}$ from the combinations of site factors arise. This per force necessitates the introduction and representation of $\boldsymbol{\alpha}(K,h)$ that involves $x$ and $y$ because external fields (via $y$) induce additional pairwise interactions (denoted by $x$) that lead to a new spin-spin interaction strength. Obviously, in the absence of magnetic field, $y=0$ and this complication can be easily surmounted.

In the subsequent analysis, it is instructive to extend the lucid analysis of Glasser [8] formulated earlier for $H\ =0$. Thus,

$$Q_{(h\geq 0)} = 2^N(1-x^2)^{-N}\,(1-y^2)^{-\frac{N}{4}}\,\exp\left(-\frac{1}{2}\boldsymbol{\alpha}(K,h)\sum_{l=1}^{\infty}S_l\,(x,y)\right) \tag{8}$$

where $S_l$ denotes the contribution of loops or sub graphs of size $l$ . It corresponds to the sum over all closed loops of length $l$ on the lattice, each loop weighted by the product of various effective edge weights along the loop. The factor $S_l$ enumerates all closed loops of length $l$ with clockwise and anticlockwise directions. Since each loop is counted twice, it is necessary to divide the sum in equation (8) by two.

In two-dimensional lattices of coordination number 4, directed edges can exist only in four directions. Hence, we consider the four directions as necessitated by the square lattice assumption and subsequently generalise to infinite sites, by invoking lattice periodicity and by recalling the elegance of Fourier transformation. As in Glasser [8], let $v$ denote the pair k$\mu$ , while the elements of $4N\times 4N$ matrix may be constructed as follows:

$$M_l(v,v') = \Sigma\text{ paths from } v' \text{ to v with } l \text{ bonds}$$
$$M_0^h(v,v') = \delta_{v,v'}$$

The matrix $M_1$ is defined in such a way that its entry $M_1(v,v')$ encodes the weight of a single step from vertex $v'$ to $v$ and this should include the edge weight between neighbouring spins. The trace of the matrix plays a central role in the subsequent analysis.

( $l^{-1}\, tr(M_1^l) = l^{-1}\sum_v M_1^l(v,v')$ sums over all length -$l$ paths that start and end at the same vertex $v$.The $l^{-1}$ term is included because while counting the loops using the trace of the matrix, the same loop is counted $l$ times.

$$M_l(v,v') = \sum_{v_1,\dots,v_{l-1}} M_l(v,v_1)M_l(v_1,v_2)\dots M_l(v_{l-1},v') = M_1^{h^l}(v,v')$$

$$S_l(x,y) = l^{-1}\sum M_l(v,v') = l^{-1}\, Tr\left\{\left(M_1^{(h)}\right)^l\right\} = l^{-1}\sum_i (\lambda_i)^l$$

$$S_l(x,y) = l^{-1}\sum_i (\lambda_i)^l \tag{9}$$

Employing the above equation, we derive the partition function as

$$Q_{(h\geq 0)} = 2^N(1-x^2)^{-N}(1-y^2)^{-\frac{N}{4}}\left\{e^{\left(-\frac{1}{2}\alpha(K,h)\sum_i\sum_{l=1}^{\infty}\frac{(\lambda_i)^l}{l}\right)}\right\}$$

$$Q_{(h\geq 0)} = 2^N(1-x^2)^{-N}(1-y^2)^{-\frac{N}{4}}\left\{\left(\alpha(K,h)\prod_i(1-\lambda_i)\right)^{\frac{1}{2}}\right\} \tag{10}$$

In order to keep our analysis similar to that leading to the exact solution, we have kept the function $\alpha(K,h)$ isolated from the determinant at this stage. It is possible to eliminate this strategy and we will revert to this approach in subsequent studies.

At this stage, it is essential to indicate the directions associated with each site in the square lattice for tracking the path. In general, the weight associated with a vertex during its path is assigned $e^{\frac{i\theta}{2}}$. Consequently, the path weights turn out to be $i^{\frac{1}{2}}$, $i^{-\frac{1}{2}}$ and 1. These weights are assigned such that $e^{\frac{i\pi}{4}} = i^{\frac{1}{2}}$ is for the clockwise turn , $e^{-\frac{i\pi}{4}} = i^{-\frac{1}{2}}$ pertains to anti-clockwise turn and $e^{-\frac{i0}{2}} = 1$ indicates a straight path.

Furthermore, $\beta\,(K,h)$ denotes the matrix element (weight) for steps between sites in the presence of both $x$ and $y$. When $y \neq 0$, the hitherto-known matrix *vis a vis* path expansion will be unable to estimate the true statistical weights. The function $\beta\,(K,h)$ encompasses $x$ as well as $y$ and it becomes equal to $x$ (= tanh $K$ ) when $y = 0$.

To interpret the four elements of $M_1$ matrix, we designate the lattice sites in the two directions as $k_1, k_2, k_1', k_2'$ respectively and $\delta(a-b)$ denotes the Kronecker delta.

$$M_1(k_1,v') = \beta\,(K,h)\Big\{[\delta(k_1-k_1')\,\delta(k_2-k_2'-1)\,]\,\delta_{\mu',1} + i^{-\frac{1}{2}}\delta(k_1-k_1'+1)\,\delta(k_2-k_2')\delta_{\mu',2} + i^{\frac{1}{2}}\delta(k_1-k_1'-1)\,\delta(k_2-k_2')\delta_{\mu',4}\Big\}$$

$$M_1(k_2,v') = \beta\,(K,h)\,\Big\{\delta(k_1-k_1'+1)\,\delta(k_2-k_2')\,\delta_{\mu',2} + i^{\frac{1}{2}}\delta(k_1-k_1')\,\delta(k_2-k_2'-1)\delta_{\mu',1} + i^{-\frac{1}{2}}\delta(k_1-k_1')\,\delta(k_2-k_2'+1)\delta_{\mu',3}\Big\}$$

$$M_1(k_3, v') = \beta\,(K,h)\Big\{\, i^{\frac{1}{2}}\delta(k_1 - k_1' + 1)\,\delta(k_2 - k_2')\delta_{\mu',2} + \delta(k_1 - k_1')\,\delta(k_2 - k_2' + 1)\ \ \delta_{\mu',3} + i^{-\frac{1}{2}}\delta(k_1 - k_1' - 1)\,\delta(k_2 - k_2')\delta_{\mu',4}\Big\} \tag{11}$$

$$M_1(k_4, v') = \beta\,(K,h)\Big\{\, i^{-\frac{1}{2}}\delta(k_1 - k_1')\,\delta(k_2 - k_2' - 1)\delta_{\mu',1} + i^{\frac{1}{2}}\delta(k_1 - k_1')\,\delta(k_2 - k_2' + 1)\ \ \delta_{\mu',3} + \delta(k_1 - k_1' - 1)\,\delta(k_2 - k_2')\delta_{\mu',4}\Big\}$$

As mentioned earlier, each matrix element involves the same function $(K, h)$ . This can be attributed to the fact that local statistical weight for any closed loops is identical everywhere, depending only on the values of the nearest neighbour interaction and magnetic field, and not on the site indices.

A schematic representation of the $M_1$ matrix is provided in Figure 1 , without $\beta\ (K, h)$.

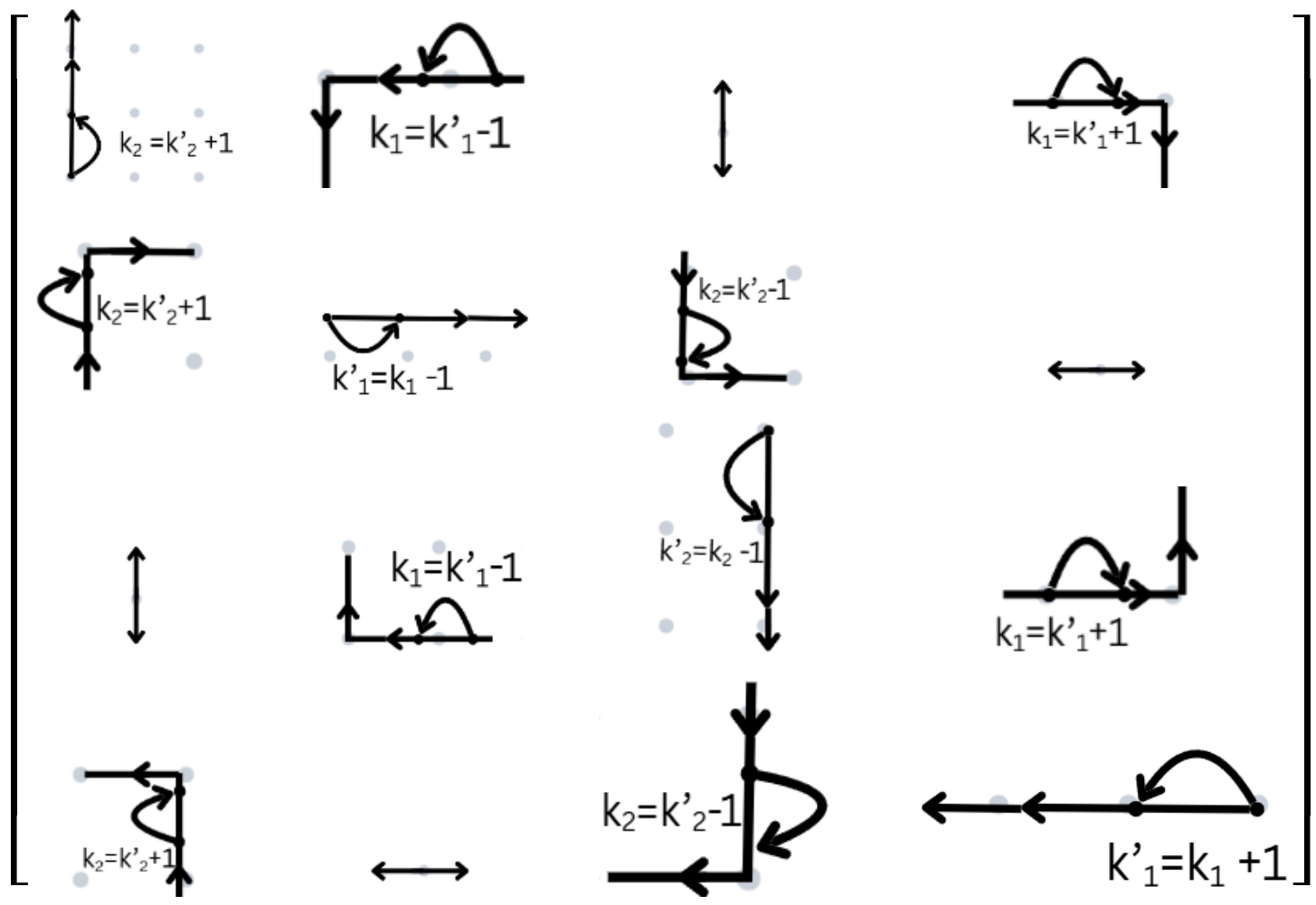

Figure 1: A schematic depiction of various matrix elements represented by equations (11)

Figure 1 illustrates the weights assigned to the Transition Matrix for a path segment moving between two directed edges.

***Matrix M1***

$$= \boldsymbol{\beta(K,h)}\begin{bmatrix} \delta(k_1 - k_1')\,\delta(k_2 - k_2' - 1)\,\delta_{\mu',1} & i^{-\frac{1}{2}}\delta(k_1 - k_1' + 1)\,\delta(k_2 - k_2')\delta_{\mu',2} & 0 & i^{\frac{1}{2}}\delta(k_1 - k_1' - 1)\,\delta(k_2 - k_2')\delta_{\mu',4} \\ i^{\frac{1}{2}}\delta(k_1 - k_1')\,\delta(k_2 - k_2' - 1)\delta_{\mu',1} & \delta(k_1 - k_1' + 1)\,\delta(k_2 - k_2')\ \delta_{\mu',2} & i^{-\frac{1}{2}}\delta(k_1 - k_1')\,\delta(k_2 - k_2' + 1)\delta_{\mu',3} & 0 \\ 0 & i^{\frac{1}{2}}\delta(k_1 - k_1' + 1)\,\delta(k_2 - k_2')\delta_{\mu',2} & \delta(k_1 - k_1')\,\delta(k_2 - k_2' + 1)\ \delta_{\mu',3} & i^{-\frac{1}{2}}\delta(k_1 - k_1' - 1)\,\delta(k_2 - k_2')\delta_{\mu',4} \\ i^{-\frac{1}{2}}\delta(k_1 - k_1')\,\delta(k_2 - k_2' - 1)\delta_{\mu',1} & 0 & i^{\frac{1}{2}}\delta(k_1 - k_1')\,\delta(k_2 - k_2' + 1)\ \delta_{\mu',3} & \delta(k_1 - k_1' - 1)\,\delta(k_2 - k_2')\delta_{\mu',4} \end{bmatrix}$$

We invoke the lattice periodicity in conjunction with the property of Kronecker delta and upon carrying out the discrete Fourier transformation of the matrix $M_1$, with the help of the unitary matrix, we obtain

$$T(q_1,q_2)=\beta(K,h)\begin{bmatrix} e^{-iq_2} & \frac{e^{iq_1}}{i^{\frac{1}{2}}} & 0 & e^{-iq_1}i^{\frac{1}{2}} \\ e^{-iq_2}\,i^{\frac{1}{2}} & e^{iq_1} & \frac{e^{iq_2}}{i^{\frac{1}{2}}} & 0 \\ 0 & e^{iq_1}i^{\frac{1}{2}} & e^{iq_2} & \frac{e^{-iq_1}}{i^{\frac{1}{2}}} \\ \frac{e^{-iq_2}}{i^{\frac{1}{2}}} & 0 & e^{iq_2}i^{\frac{1}{2}} & e^{-iq_1} \end{bmatrix} \tag{12}$$

Furthermore, determinant of the above matrix can be written as $\det\left|\left(I-T(q_1,q_2)\right)\right| = (1-\lambda_1)(1-\lambda_2)(1-\lambda_3)(1-\lambda_4)$. The eigenvalues arising from the above matrix are shown in the Appendix A. Using the definition of the trace of the matrix, we deduce $Q$ as

$$Q_{(h\geq 0)} = 2^N\,(1-x^2)^{-N}(1-y^2)^{-\frac{N}{2}}\left\{\prod_q \boldsymbol{\alpha(K,h)}\ \det\left|\left(I-T(q_1,q_2)\right)\right|\right\}^{\frac{1}{2}} \tag{13}$$

The product of $(I-\lambda_i)$ for each block of the 4N×4N matrix follows as

$$\det\left|\left(I-T(q_1,q_2)\right)\right| = (1+\beta(K,h)^2)^2 - 2\beta(K,h)(1-\beta(K,h)^2)(cosq_1+cosq_2)$$

$$\alpha(x,y)\det\left|\left(I-T(q_1,q_2)\right)\right| \\ = \alpha(K,h)(1+\beta(K,h)^2)^2 - 2\alpha(K,h)\beta(K,h)(1-\beta(K,h)^2)(cosq_1+cosq_2) \tag{14}$$

Employing equation (14) in equation (13), we obtain

$$Q_{(h\geq 0)} = 2^N(1-x^2)^{-N}(1-y^2)^{-\frac{N}{4}} \\ \left\{\prod_q [\alpha(K,h)(1+\beta(K,h)^2)^2 - 2\alpha(K,h)\beta(K,h)(1-\beta(K,h)^2)(cosq_1+cosq_2)]\right\}^{\frac{1}{2}} \tag{15}$$

Since our primary interest is in the thermodynamic limit of $N\ \rightarrow\ \infty$, it is advantageous to convert the above summation into a double integral involving $q_1$ and $q_2$ as

$$\frac{1}{N}\,lnQ_{(h\geq 0)} = ln(2) - ln(1-x^2) - \frac{1}{4}ln(1-y^2) \\ +\frac{1}{8\pi^2}\int_0^{2\pi}\int_0^{2\pi} ln\{\alpha(K,h)(1+\beta(K,h)^2)^2 - 2\alpha(K,h)\beta(K,h)(1-\beta(K,h)^2)(cosq_1+cosq_2)\}\,dq_1\,dq_2 \tag{16}$$

While equation (16) is analogous to the usual integral arising in Onsager's exact solution, a more appropriate procedure will be to carry out the inversion of ln Q and deduce the functions $\alpha(K,h)$ *and* $\beta(K,h)$ therefrom in a systematic manner.

In a fortuitous manner, the series expansions for $Q\ (h,K)$ are available at low and high temperatures in various formats, the most convenient being the equations (1.8.10) and (1.8.11) reported by Baxter [31] on account of the anticipated invariance of the partition function with respect to the sign of $h$. When the magnetic field is zero, the above equation becomes identical with equation (19) of Glasser [8]. The above methodology is identical for low and high temperatures at this stage. The functions $\alpha(K,h)$ and $\beta\ (K,h)$ however cannot be identical on account of the inherent differences in the coupling between the magnetic field and interaction energies.

**Partition function and magnetization at low temperatures**

$$\frac{1}{N}\ lnQ_{(h\geq 0)} = ln(2) + \frac{1}{2} ln\left(\frac{\alpha(K,h)(1+\beta(K,h)^2)^2}{(1-x^2)^2(1-y^2)^{\frac{1}{2}}}\right) + \frac{1}{8\pi^2}\int_0^{2\pi}\int_0^{2\pi} ln\left\{1 - \frac{2\beta(K,h)(1-\beta(K,h)^2)}{(1+\beta(K,h)^2)^2}(cosq_1 + cosq_2)\right\} dq_1\, dq_2 \qquad (17)$$

The above integral can be further simplified as shown in Appendix B and leads to the partition function as

$$\frac{1}{N}\ lnQ_{(h\geq 0)} = \frac{1}{2}\ln\left(\frac{2\alpha(K,h)(1+\beta(K,h)^2)^2}{(1-x^2)^2(1-y^2)^{\frac{1}{2}}}\right) + \frac{1}{\pi}\int_0^{\frac{\pi}{2}} ln\left(1 + \sqrt{1 - \left(\frac{4\beta(K,h)(1-\beta(K,h)^2)}{(1+\beta(K,h)^2)^2}\right)^2 sin^2\varphi}\right) d\varphi \qquad (18)$$

Here we assume the following function for $\alpha(K,H)$ and $\beta(K,H$ for $h \geq 0$

$$\alpha(K,h) = \frac{g(u,h)^2}{4} sech(h)sech^4(K)\left[f(u,h)^{\frac{1}{4}}\left(4f(u,h)^{\frac{1}{2}} + 2cosh(2h+4K) - 2\right)^{\frac{1}{2}} + 2f(u,h)^{\frac{1}{2}} + sinh^2(h+2K)\right]$$

$$\beta(K,h) = \frac{1}{2sinh(h+2K)}\left[-2f(u,h)^{\frac{1}{4}} + \left(-2\ + 4f(u,h)^{\frac{1}{2}} + 2cosh(2h+4K)\right)^{\frac{1}{2}}\right]$$

where $f(u,H)$ is a truncated polynomial and is given by

$$f(u,h)_{h\geq 0} = e^{2h} + u^4(-12 + 8e^{-2h} + 4e^{-4h}) + u^6(-56e^{-2h} + 20e^{-4h} + 24e^{-6h} + 12e^{-8h}) + u^8(129e^{-2h} - 360e^{-4h} - 36e^{-6\,h} + 104e^{-8h} + 106e^{-10h} + 48e^{-12h} + 9e^{-14h}) \ldots \qquad (19)$$

with u being the conventional low-temperature variable in the series expansion, defined as $u = e^{-2K}$

$$g(u,h)_{h\geq 0} = 1 - u^2(e^{-h} + e^{-2h}) + u^4(2e^{-2h} - 3e^{-3h} + e^{-4h}) + u^6(3e^{-3h} + 8e^{-4h} - 10e^{-5h} + e^{-6h} - 2e^{-7h}) + u^8(-9e^{-4h} + 27e^{-5h} + 32e^{-6h} - 32e^{-7h} + 6e^{-8h} - 18e^{-9h} - 6e^{-11h}) + u^{10}\left(-\frac{31}{2}e^{-5h} - 89e^{-6h} + 174e^{-7h} + 122e^{-8h} - 70e^{-9h} + 54e^{-10h} - 92e^{-11h} + 14e^{-12h} - 69e^{-13h} - 24e^{-15h} - \frac{9}{2}e^{-17h}\right) + \cdots \qquad (20)$$

Hence

$$\frac{1}{N}lnQ_{h\geq 0} = ln(g(u,h)_{h\geq 0}) + ln\left\{2^{\frac{1}{2}}\left(f(u,h)_{h\geq 0}^{\frac{1}{2}} + sinh^2(h+2K)\right)^{\frac{1}{2}}\right\}$$
$$+\frac{1}{\pi}\int_0^{\frac{\pi}{2}} ln\left(1+\sqrt{1-\kappa(u,h)_{h\geq 0}^2 sin^2\varphi}\right)d\varphi \qquad (21)$$

As in the Onsager's exact solution for the zero-field partition function, we define a new parameter $\kappa(K,h)$ as

$$\kappa(K,h)_{\boldsymbol{h\geq 0}} = 2\frac{f(u,h)_{h\geq 0}^{\frac{1}{4}} sinh(2K+h)}{f(u,h)_{h\geq 0}^{\frac{1}{2}} + sinh^2(2K+h)} \qquad (22)$$

$f(u,h)$ becomes unity at h =0 and hence leads to the corresponding equation (4) for $\kappa$. Thus equation (21) implies that a straight-forward generalization of Onsager's exact solution does exist, although further improvisation is still required.

The partition function deduced earlier (21) is valid for $h \geq 0$ **.**When $h \leq 0$, all the terms involving $h$ needs to be changed as $-h$ .Consequently, the total partition function for positive and negative values of $h$ can be formulated as suggested in [31].

$$Q(K,h) = \frac{1}{2}\left\{\begin{array}{l} e^{N\left[ln(g(u,h)_{h\geq 0})+ln\left\{2^{\frac{1}{2}}\left(f(u,h)_{h\geq 0}^{\frac{1}{2}}+sinh^2(h+2K)\right)^{\frac{1}{2}}\right\}+\frac{1}{\pi}\int_0^{\frac{\pi}{2}} ln\left(1+\sqrt{1-\kappa(u,h)_{h\geq 0}^2 sin^2\varphi}\right)d\varphi\right]} \\ + e^{N\left[ln(g(u,h)_{h\leq 0})+ln\left\{2^{\frac{1}{2}}\left(f(u,h)_{h\leq 0}^{\frac{1}{2}}+sinh^2(h+2K)\right)^{\frac{1}{2}}\right\}+\frac{1}{\pi}\int_0^{\frac{\pi}{2}} ln\left(1+\sqrt{1-\kappa(u,h)_{h\leq 0}^2 sin^2\varphi}\right)d\varphi\right]} \end{array}\right\} \qquad (23)$$

It is easy to infer that Q is an even function of the magnetic field. The equations for $f(u,h)_{h\geq 0}$ and $g(u,h)_{h\geq 0}$ have been given in (19) and (20) respectively and when $h$ is negative, the sign of $h$ in those equations should be reversed. For $h \leq 0,\ \kappa$ is given by

$$\kappa(K,h)_{h\leq 0} = 2\frac{f(u,h)_{h\leq 0}^{\frac{1}{4}} sinh(2K-h)}{f(u,h)_{h\leq 0}^{\frac{1}{2}} + sinh^2(2K-h)}.$$

*in lieu* of equation (22). An equivalent expression for equation (3) in terms of Gauss hypergeometric series has been derived by Viswanathan [32] exploiting the isomorphism with Mahler measures. The same expression is also obtainable [33] by analytically evaluating the integral of equation (3). In the present analysis when $h \neq 0$, the partition function can also be represented in an equivalent manner as

$$\frac{1}{N}lnQ_{h\geq 0} = ln[2\,cosh(2K+h)\ g(u,h)_{h\geq 0}] - \frac{\kappa(u,h)_{h\geq 0}^2}{16}\ {}_4F_3\left(1,1,\frac{3}{2},\frac{3}{2},2,2,2,\kappa(u,h)_{h\geq 0}^2\right) \qquad (24)$$

where $\kappa(K,h)_{\boldsymbol{h\geq 0}}$ is given by equation (22). The interchange of u and K is necessitated since the preferred variable for low temperature series expansion for thermodynamic functions is $u\ (= e^{-2K})$ . The above equation represents the partition function of the two-dimensional nearest neighbour Ising model when the field is non-zero and is entirely new.

At $h = 0, g(u,0) = f(u,0) = 1$ and hence the partition function becomes

$$\frac{1}{N} lnQ_{h=0} = ln[2\,cosh(2K)\,] - \frac{\kappa^2}{16}\ {}_4F_3\left(1,1,\frac{3}{2},\frac{3}{2},2,2,2,\kappa^2\right) \tag{25}$$

identical with the equation first deduced by Viswanathan [32].

**Magnetization when $\frac{T}{T_c} \le 1$**

Employing the customary definition of magnetization ($M$) as the first derivative of the Helmholtz free energy with respect to the magnetic field [28], we derive the magnetization using equation (21):

$$M = \left[1 - \frac{f(u,h)_{h\geq 0}}{\sinh^4(2K+h)}\right]^{\frac{1}{8}} \tag{26}$$

where $f(u,h)_{h\geq 0}$ is given by equation (19). Upon substituting the function $f(u,h)$ given earlier, we obtain the series expansion for the field-dependent magnetization as

$$M\,(h,K) = 1 - 2u^4e^{-2h} - 8u^6e^{-4h} + u^8(10e^{-4h} - 36e^{-6h} - 8e^{-8h}) + u^{10}(96e^{-6h} - 144e^{-8h} - 80e^{-10h} - 24e^{-12h}) + u^{12}(-62\,e^{-6h} + 680e^{-8h} - 430e^{-10h} - 480e^{-12h} - 308e^{-14h} - 96e^{-16h} - 18e^{-18h}) + \cdots$$

The above equation is identical with the known series expansion for low temperature magnetization series, till $u^{12}$ term [34]. We reiterate that the above equation arises solely from the field-dependent partition function.

**Partition function and magnetization at high temperature**

Equation (5) for the spontaneous magnetization at $\frac{T}{T_c} \le 1$ was deduced by Yang [27] and is a convenient starting point to extend the same for $h \neq 0$. However, since the spontaneous magnetization is zero when $\frac{T}{T_c} > 1$, there is no *a priori* method of verifying the correctness of $M$ equation at $h \neq 0$. This renders the analysis pertaining to $h \neq 0$ and $\frac{T}{T_c} > 1$, far more challenging than that for $\frac{T}{T_c} < 1$, analysed earlier.

In view of this, we postulate a hierarchy of equations for $\alpha(K,h)$, keeping in mind, the corresponding terms in the series expansions pertaining to the zero-field susceptibility.

When $\frac{T}{T_c} > 1$, equation (18) is modified by assuming that $\alpha(K,h) = g_1(h,K)^2(1-y^2)^{\frac{1}{2}}$; furthermore, when $h = 0, \beta(K) = \tanh K = x$. Consequently,

$$\frac{1}{N}\,lnQ_{(h\geq 0)} = \frac{1}{2}\ln\left(\frac{2\alpha(K,h)(1+\beta(K,h)^2)^2}{(1-x^2)^2(1-y^2)^{\frac{1}{2}}}\right) + \frac{1}{\pi}\int_0^{\frac{\pi}{2}} ln\left(1+\sqrt{1-\left(\frac{4\beta(K,h)(1-\beta(K,h)^2)}{(1+\beta(K,h)^2)^2}\right)^2 sin^2\varphi}\right)d\varphi$$

$$\frac{1}{N}\, lnQ_{(h\geq 0)} = \frac{1}{2}\ln\left(\frac{2g_1(h,K)^2(1-y^2)^{\frac{1}{2}}(1+x^2)^2}{(1-x^2)^2(1-y^2)^{\frac{1}{2}}}\right) + \frac{1}{\pi}\int_0^{\frac{\pi}{2}} ln\left(1+\sqrt{1-\left(\frac{4x(1-x^2)}{(1+x^2)^2}\right)^2 sin^2\varphi}\right)d\varphi$$

$$\frac{1}{N}\, lnQ_{(h\geq 0)} = \frac{1}{2}\ln\left(\frac{2g_1(h,K)^2(1+x^2)^2}{(1-x^2)^2}\right) + \frac{1}{\pi}\int_0^{\frac{\pi}{2}} ln\left(1+\sqrt{1-\left(\frac{4x(1-x^2)}{(1+x^2)^2}\right)^2 sin^2\varphi}\right)d\varphi$$

$$\frac{1}{N}\, lnQ_{(h\geq 0)} = ln\left[2^{\frac{1}{2}} cosh(2K)\, g_1(h,K)\right] + \frac{1}{\pi}\int_0^{\frac{\pi}{2}} ln\left(1+\sqrt{(1-\kappa(x)^2 sin^2\phi)}\right)d\phi \qquad (27)$$

where, $\kappa(x) = \frac{4\,x\,(1-x^2)}{(1+x^2)^2}$ which is identical with the dependence of $\kappa$ at $h = 0$ (equation 4). This suggests that a modification of $\kappa$ for non-zero values of h is required essentially for $\boldsymbol{\frac{T}{T_c} \leq 1}$.

In order to deduce the magnetic equation of state, the complete dependence of M on h is required; however, to obtain the partial series coefficients of zero-field susceptibility, terms higher than $\tanh^2$ (h) can be neglected as shown in Table 1b.

*At the first level of approximation*, we assume that

$$g_1(h,K) = sinh(h)\, tanh(h)\, tanh(2K) + cosh(h) \qquad (28)$$

Hence

$$\frac{1}{N} lnQ = ln\left[\, 2^{\frac{1}{2}} cosh(h)\, cosh(2K) + 2^{\frac{1}{2}} sinh(h)\, tanh(h)\, sinh(2K)\,\right] + \frac{1}{\pi}\int_0^{\frac{\pi}{2}} ln\left(1+\sqrt{(1-\kappa(x)^2 sin^2\phi)}\right)d\phi \qquad (29)$$

At $K = 0$, the above equation leads to

$$\frac{1}{N} lnQ = ln(2cosh(h))$$

as anticipated. At $h = 0$, equation (29) leads to equation (3). Interestingly, the field-dependent magnetization from equation (27) is

$$M = \frac{g_1'(h,K)}{g_1(h,K)}$$

The field-dependent magnetization follows as

$$M = \frac{tanh(h)\left(1 + 4tanh(K) + tanh^2(K) - 2\, tanh(K)tanh^2(h)\right)}{1 + tanh^2(K) + 2tanh(K)\, tanh^2(h)} \qquad (30)$$

if equation (28) is employed for $g_1(h, K)$, thereby yielding the first two terms of the zero-field susceptibility series ( $\chi_{h=0}$) series viz

$$\left(\frac{dM}{dh}\right)_{h=0} = 1 + 4x$$

In order to obtain various terms in the series expansion of zero field susceptibility, $g_1(h, K)$ needs to be suitably modified as shown in Table 1b.

**Table 1a: A hierarchy of functions for $g_1(K, h)$ to deduce the dependence of the magnetization on magnetic field and interaction energy**

| **Function for $g_1(K, h)$** | $M$ |
|---|---|
| $sinh(h)\ tanh(h)\ tanh(2K) + cosh(h)$ | $\frac{tanh(h)\begin{pmatrix}1 + 4\,tanh(K) + tanh^2(K) \\ -2\,tanh(K)tanh^2(h)\end{pmatrix}}{1 + tanh^2(K) + 2tanh(K)\ tanh^2(h)}$ |
| $\frac{\begin{bmatrix} cosh(h)(tanh(K) + 1)^4 \\ - sech(h)tanh(K)\begin{pmatrix}5\,sech^2(h)tanh(K) \\ + 2\,tanh^2(K) + 2\end{pmatrix}\end{bmatrix}}{tanh(K)(tanh(K) + 2)(tanh^2(K) + 1) + 1}$ | $\frac{tanh(h)\begin{bmatrix}15x^2tanh^4(h) - \\ 2xtanh^2(h)(x^2 + 15x + 1) + \\ x^4 + 6x^3 + 21x^2 + 6x + 1\end{bmatrix}}{\begin{bmatrix}-5x^2tanh^4(h) + \\ 2xtanh^2(h)(x^2 + 5x + 1) + \\ x^4 + 2x^3 + x^2 + 2x + 1\end{bmatrix}}$ |
| $\frac{e^h\begin{bmatrix}-26\,sech^6(h)tanh^3(K) \\ - 5\,sech^4(h)tanh^4(K) \\ + 14\,sech^4(h)tanh^3(K) \\ - 5\,sech^4(h)tanh^2(K) \\ - 2\,sech^2(h)tanh^5(K) \\ - 4\,sech^2(h)tanh^4(K) \\ - 4\,sech^2(h)tanh^3(K) \\ - 4\,sech^2(h)tanh^2(K) \\ - 2\,sech^2(h)tanh(K) \\ + tanh^6(K) + 6\,tanh^5(K) \\ + 15\,tanh^4(K) + 20\,tanh^3(K) \\ + 15\,tanh^2(K) + 6\,tanh(K) + 1\end{bmatrix}}{\begin{bmatrix}\frac{1}{16}sech^6(K)\begin{pmatrix}14\sinh(2K) + \\ 6\sinh(6K) \\ + 9\cosh(2K) \\ + 7\,cosh(6K)\end{pmatrix}\end{bmatrix}(tanh(h) + 1)}$ | $\frac{tanh(h)\begin{pmatrix}-130\,x^3tanh^6(h) + 15\,x^4tanh^4(h) \\ + 348x^3tanh^4(h) + 15x^2tanh^4(h) \\ -2x^5tanh^2(h) - 34x^4tanh^2(h) \\ -310x^3tanh^2(h) - 34x^2tanh^2(h) \\ -2xtanh^2(h) + x^6 \\ +8x^5 + 34x^4 \\ +112x^3 + 34x^2 \\ + 8x + 1\end{pmatrix}}{\begin{pmatrix}26x^3tanh^6(h) - 5x^4tanh^4(h) \\ -64x^3tanh^4(h) - 5x^2tanh^4(h) \\ +2x^5tanh^2(h) + 14x^4tanh^2(h) \\ +54x^3tanh^2(h) + 14x^2tanh^2(h) \\ +2xtanh^2(h) + x^6 \\ +4x^5 + 6x^4 \\ + 4x^3 + 6x^2 \\ +4x + 1\end{pmatrix}}$ |
| $\dfrac{\left\{\begin{matrix} tanh^8(K) + 6\,tanh^7(K) \\ + 15\,tanh^6(K) + 20\,tanh^5(K) \\ +16\,tanh^4(K) + 20\,tanh^3(K) \\ -172x^4\,tanh^8(h) + \\ 2\,tanh^3(K)\begin{bmatrix} 13\,tanh^2(K) + \\ 270\,tanh(K) \\ +13 \end{bmatrix} tanh^6(h) - \\ tanh^2(K)\,tanh^4(h)\begin{bmatrix} 5\,tanh^4(K) + \\ 74\,tanh^3(K) + \\ 602\,tanh^2(K) + \\ 74\,tanh(K) + 5 \end{bmatrix} + \\ 2\,tanh(K)\,tanh^2(h)\begin{bmatrix} tanh^6(K) + \\ 9\,tanh^5(K) + \\ 42\,tanh^4(K) + \\ 144\,tanh^3(K) + \\ 42\,tanh^2(K) + \\ 9tanh(K) + 1 \end{bmatrix} \\ +15\,tanh^2(K) + 6tanh(K) + 1 \end{matrix}\right\} cosh(h)}{\left[\frac{1}{32}sech^8(K)\begin{pmatrix} 22\,sinh(4\,K) \\ + 13\,sinh(8\,K) \\ + 28\,cosh(4\,K) \\ + 12\,cosh(8\,K) - 8 \end{pmatrix}\right]}$ | $\dfrac{tanh(h)\begin{pmatrix} x^8 - 2x^7tanh^2(h) \\ + 10x^7 + 15x^6tanh^4(h) \\ - 38x^6tanh^2(h) + 51x^6 \\ - 130x^5tanh^6(h) + 378x^5tanh^4(h) \\ - 380x^5tanh^2(h) + 188x^5 + \\ 1204x^4tanh^8(h) \\ - 4076x^4tanh^6(h) \\ + 5046x^4tanh^4(h) \\ - 2696x^4tanh^2(h) \\ + 592x^4 - 130x^3tanh^6(h) + \\ 378x^3tanh^4(h) - 380x^3tanh^2(h) + \\ 188x^3 + 15x^2tanh^4(h) \\ - 38x^2tanh^2(h) + 51x^2 \\ -2xtanh^2(h) + 10x + 1 \end{pmatrix}}{\begin{pmatrix} x^8 + 2x^7tanh^2(h) \\ + 6x^7 - 5x^6tanh^4(h) \\ + 18x^6tanh^2(h) + 15x^6 \\ + 26x^5tanh^6(h) - 74x^5tanh^4(h) \\ + 84x^5tanh^2(h) + 20x^5 \\ - 172x^4tanh^8(h) + 540x^4tanh^6(h) \\ - 602x^4tanh^4(h) + 288x^4tanh^2(h) + \\ 16x^4 + 26x^3tanh^6(h) \\ - 74x^3tanh^4(h) + 84x^3tanh^2(h) \\ + 20x^3 - 5x^2tanh^4(h) \\ + 18x^2tanh^2(h) + 15x^2 \\ + 2xtanh^2(h) + 6x + 1 \end{pmatrix}}$ |

**Table 1b: Equation for magnetization (M) arising from the partition function to obtain partial series coefficients of the zero-field susceptibility**

| ***M*** | **The zero-field susceptibility series** |
|---|---|
| $\dfrac{tanh(h)(1 + 4\,tanh(K) + tanh^2(K))}{1 + tanh^2(K)}$ | $1 + 4x$ |
| $\dfrac{tanh(h)[x^4 + 6x^3 + 21x^2 + 6x + 1]}{[x^4 + 2x^3 + x^2 + 2x + 1]}$ | $1 + 4x + 12x^2$ |
| $\dfrac{tanh(h)\,(x^6 + 8x^5 + 34x^4 + 112x^3 + 34x^2 + 8x + 1)}{(x^6 + 4x^5 + 6x^4 + 4x^3 + 6x^2 + 4x + 1)}$ | $1 + 4x + 12x^2 + 36x^3$ |
| $\dfrac{tanh(h)\,(x^8 + 10x^7 - 38x^6 + 51x^6 + 188x^5 + 592x^4 + 188x^3 + 51x^2 + 10x + 1)}{x^8 + 6x^7 + 15x^6 + 20x^5 + 16x^4 + 20x^3 + 15x^2 + 6x + 1}$ | $1 + 4x + 12x^2 + 36x^3 + 100x^4$ |

**Table 2: Compilation of the equations for the partition function and magnetization**

| **Equation** | **Remarks** |
|---|---|
| $\frac{1}{N}lnQ_{\boldsymbol{h\geq 0}} = \ln(g(u,h)_{h\geq 0}) + \ln\left\{2^{\frac{1}{2}}\left(f(u,h)_{h\geq 0}^{\frac{1}{2}} + sinh^2(h+2K)\right)^{\frac{1}{2}}\right\} + \frac{1}{\pi}\int_0^{\frac{\pi}{2}} \ln\left(1 + \sqrt{1 - \kappa(u,h)_{h\geq 0}^2 \sin^2\varphi}\right)d\varphi$ | $\frac{T}{T_c} \leq 1; h \geq 0$ |
| $M = \left[1 - \frac{f(u,h)_{h\geq 0}}{\sinh^4(2K+h)}\right]^{\frac{1}{8}}$ | $\frac{T}{T_c} \leq 1; h \geq 0$ |
| $\frac{1}{N} lnQ = ln\left[2^{\frac{1}{2}} cosh(2K)\, g_1(h,K)\right] + \frac{1}{\pi}\int_0^{\frac{\pi}{2}} ln\left(1 + \sqrt{(1-\kappa(x)^2 \sin^2\phi)}\right) d\phi$ | $\frac{T}{T_c} > 1;$ |
| $M = \frac{tanh(h)(-2\,x\,tanh^2(h) + x^2 + 4\,x + 1)}{2\,x\,tanh^2(h) + x^2 + 1}$ | $\frac{T}{T_c} > 1;$ |

*The functions $f(u,h)_{h\geq 0}$, $g(u,h)_{h\geq 0}$ and $\kappa(u,h)_{h\geq 0}$ are defined in equations 19,20 and 22 respectively. When T/Tc > 1, the composition of $g_1(K,h)$ depends upon the number of terms required for the zero field susceptibility series as shown in Table 1.*

## Results and Discussion

The analysis of two-dimensional Ising models has been a central theme in condensed matter physics ever since the pioneering work of Kramers and Wannier[2] and Onsager[3].Among various subsequent methodologies to *re-derive* Onsager's exact solution, those involving graph-theoretical procedures are especially suitable in view of their simplicity as demonstrated by Feynman [6], Glasser[8], Vvodichenko[11] , Hurst and Green [12] and Berezin [13].The essential difference among these protocols consists in the manner in which the interactions between the nearest neighbors is incorporated within a matrix formalism; the simplicity in fact emanates from the enumeration of connectivity between edges. However, the inclusion of magnetic fields within this framework is rendered difficult on account of computational complexity as well as geometrical constraints. In other words, the bottleneck now consists in the precise counting of all the labelled subgraphs consisting of lines as well as areas associated with polygons. For lattices of large sizes, this enumeration is almost impossible, thus necessitating heuristic or intuitive considerations. Consequently, we have introduced separate empirical functions involving magnetic field and interaction energies such that they yield the (i) known exact solution of the partition function when the field is zero and (ii) appropriate power series expansions of magnetization and susceptibility in non-zero magnetic fields.

As mentioned earlier, the partition functions derived herein lead to the well-known equations pertaining to $h = 0$ and consequently, the variation of the internal energy, specific heat and entropy with $K$ is identical with the already-known patterns. Analogously, the critical exponents $\alpha$ and β are consistent with the known values. The essential pre-requisite for validating our results consists in deducing the critical exponents $\gamma$ and $\delta$ pertaining to the field-dependence of magnetization at the critical temperature and zero-field susceptibility respectively.

Before we describe the variation of different thermodynamic quantities on the magnetic field, it is instructive to analyse the dependence of $\kappa$ on $h$ and $K$. This is essential since $\kappa$ =1 in equation (4) yields the exact critical temperature (Figure 2a). A minor modification of equation (22) when $h > 0$ is adequate:

$$\kappa(K,h)_{h\geq 0} = 2\frac{f(u,h)_{h\geq 0}^{\frac{1}{4}} sinh(2K+h)}{f(u,h)_{h\geq 0}^{\frac{1}{2}} + sinh^2(2K+h)}$$

For $h < 0$, the above equation is still valid with a change in the sign of h.

Furthermore, the equation relating $M(h,K)$ and $\kappa(K,h)_{h\geq 0}$ is as follows:

$$\kappa(K,h) = \frac{2(1-M^8)^{\frac{1}{4}}}{1+(1-M^8)^{\frac{1}{2}}}$$

The above equation is reminiscent of equation (4), pertaining to $h = 0$. Figure 2b depicts the dependence of $\kappa$ on $h$ and $K$.

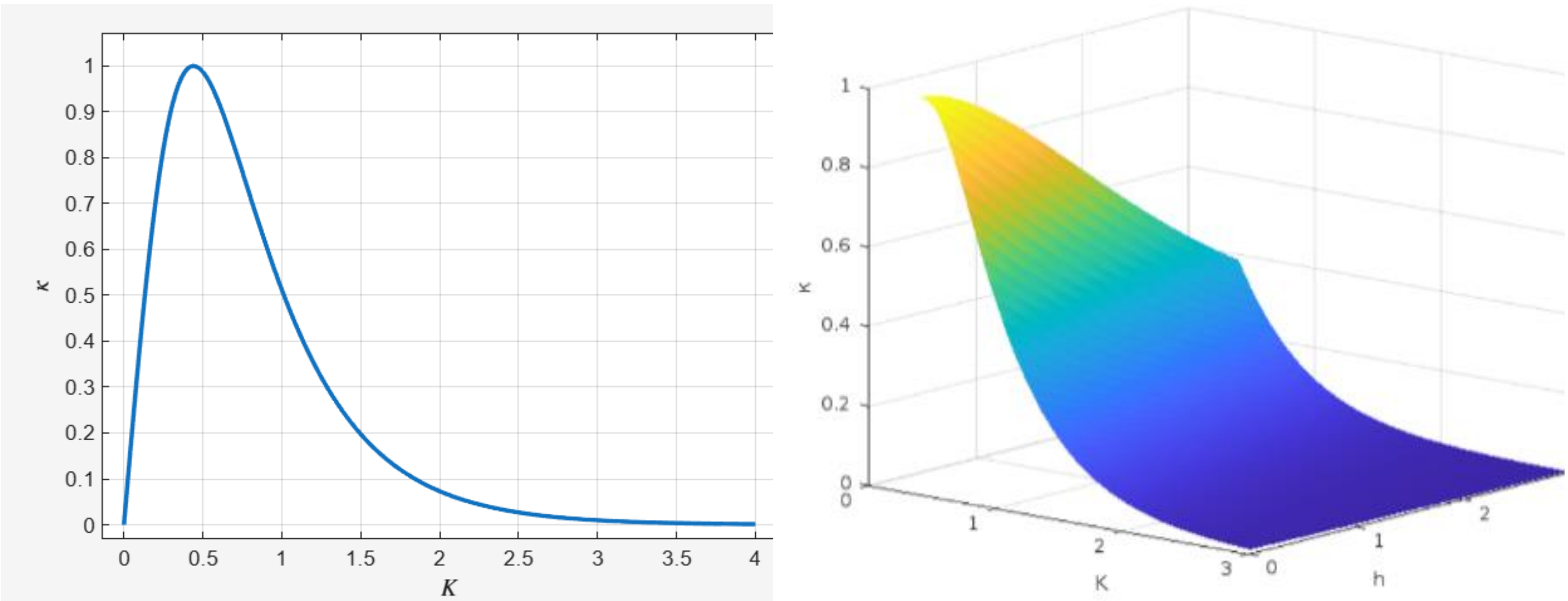

(a) (b)

Figure 2 : Dependence of $\kappa$ on the magnetic field and interaction energies for T/Tc ≤ 1 :

(a) $h = 0$ and (b) $h \neq 0$

### Partition function per site

Equations (21) and (27) provide the partition functions for various ranges of $h$ and $K$. At $H = 0$, these equations seamlessly lead to Onsager's exact solution. The straight-forward extension of equation (3) to $h \neq 0$ is indeed unexpected. Furthermore, the partition function is an even function for all values of $h$, both at high and low temperatures. The empirical functions chosen have led to correct coefficients till $u^{12}$ for the low temperature magnetization series and till $x^4$ in the high temperature susceptibility series. These were employed in all the computations herein.

The formulation of the partition function at high temperatures is rendered more difficult and a hierarchy of equations in this case can be obtained, depending upon the choice of $g_1(h,K)$. It is instructive at this stage to analyse the dependence of the partition function on $h$ and $K$ for low (Figure 3a) and high (Figure 3b) temperatures as well as for both ranges of $\frac{T}{T_c}$ ( Figure 3c). The function $g_1(h, K)$shown in the third row of Table 1 was used for all the calculations pertaining to high temperatures.

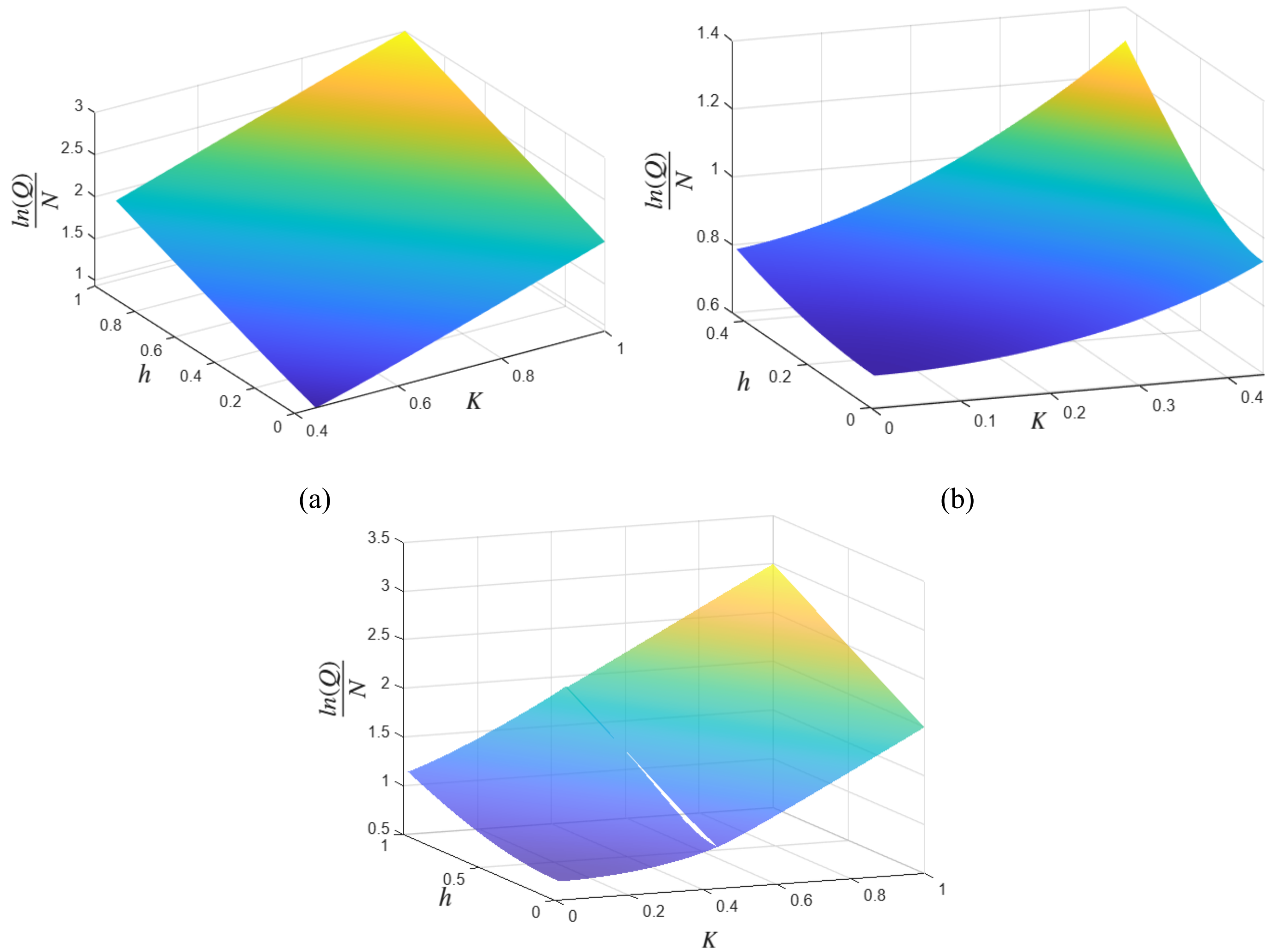

(c)

Figure 3: Dependence of the partition function per site on the magnetic field and interaction energies: (a) $\frac{T}{T_c} \leq 1$ (b) $\frac{T}{T_c} > 1$ and (c) both ranges of $\frac{T}{T_c}$

**Internal Energy**

The definition of the internal energy from the partition function is given by

$$\frac{U}{JN} = \frac{kT^2}{J}\frac{dlnQ}{dT}$$

The variation of the internal energy with magnetic field and interaction energies arising from equation (21) and (28) is shown in Figure 4a, 4b and 4c. The corresponding algebraic expression for $U(h, K)$ is shown in Appendix C

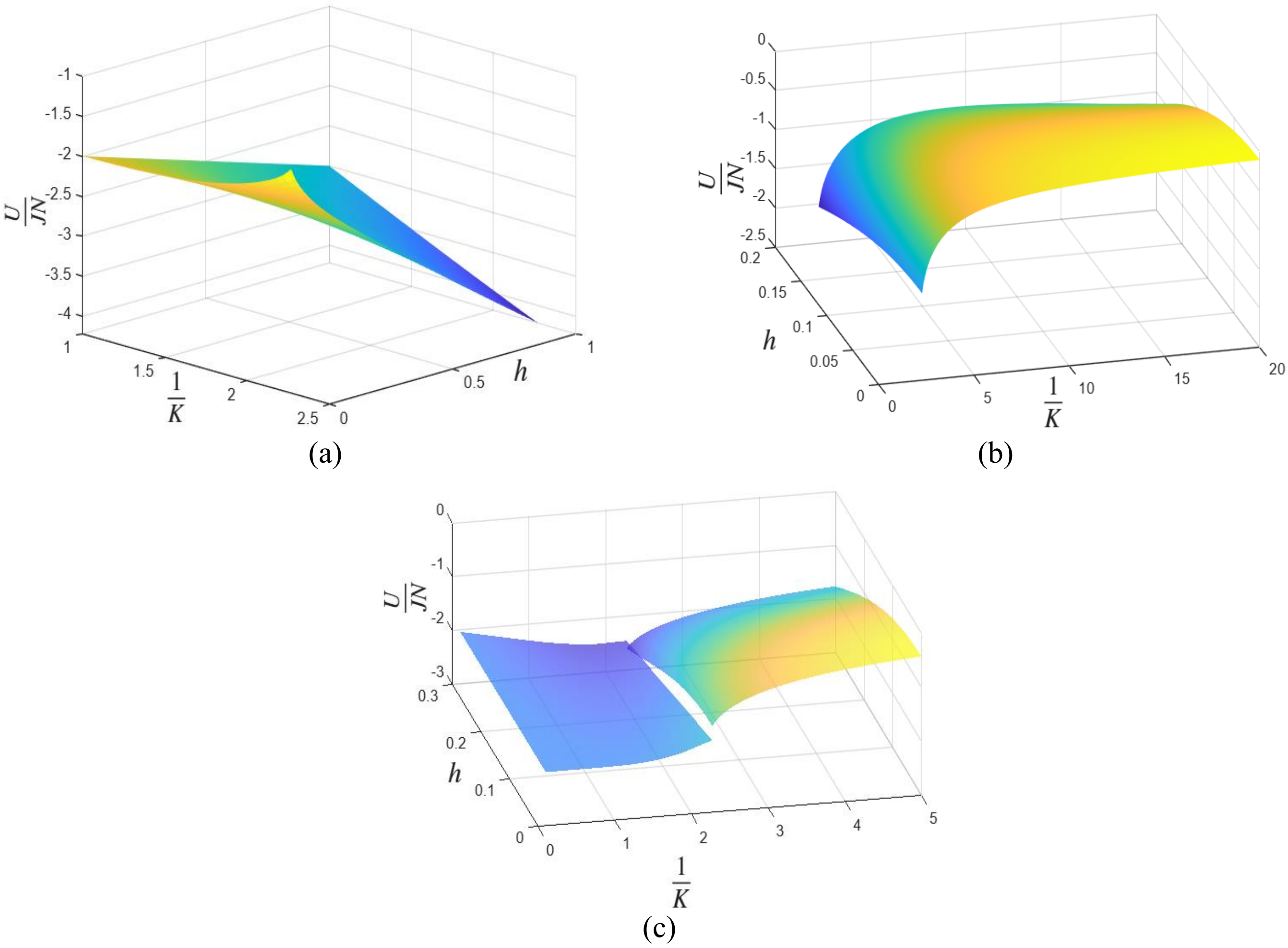

Figure 4: Dependence of the internal energy on the magnetic field and interaction energies: (a) $\frac{T}{T_c} \leq 1$ and (b) $\frac{T}{T_c} > 1$ and (c) both ranges of $\frac{T}{T_c}$

**Specific heat**

The variation of the specific heat can be easily deduced from the internal energy as

$$\frac{C_V}{Nk} = \frac{\partial}{\partial T}\left(kT^2 \frac{d(lnQ)}{dT}\right)$$

While the algebraic equation for $C_v$ is shown in Appendix C, Figure 5 depicts the variation of $C_v$ with magnetic field and interaction energies.

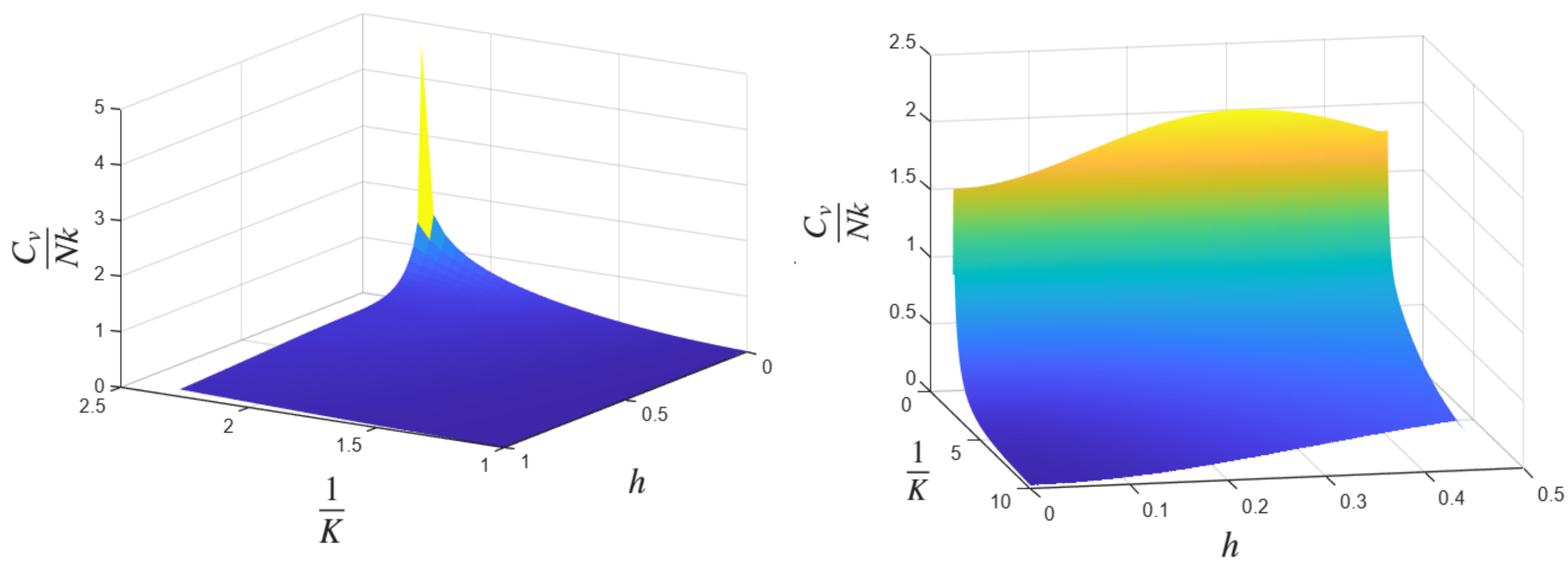

(a) (b)

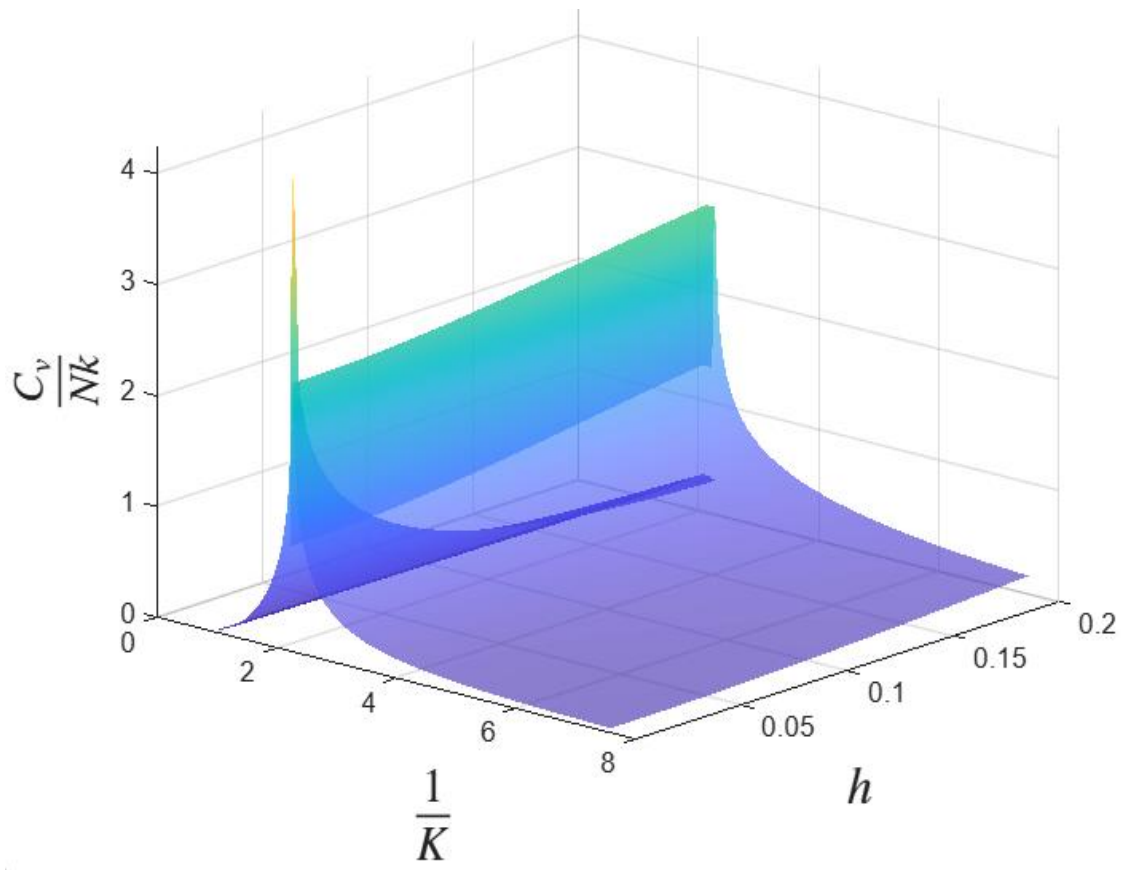

(c)

Figure 5: Dependence of the specific heat on the magnetic field and interaction energies:

(a) $\frac{T}{T_c} \leq 1$ and (b) $\frac{T}{T_c} > 1$ and (c) both ranges of $\frac{T}{T_c}$

**Entropy**

The equation for entropy (S) follows from the well-known thermodynamic equation

$F = U - TS$

Upon employing the earlier expressions for $F$ and $U$, we obtain the dependence of $S$ as shown in Appendix C. Figure 6 depicts the dependence of entropy on the magnetic field and interaction energies.

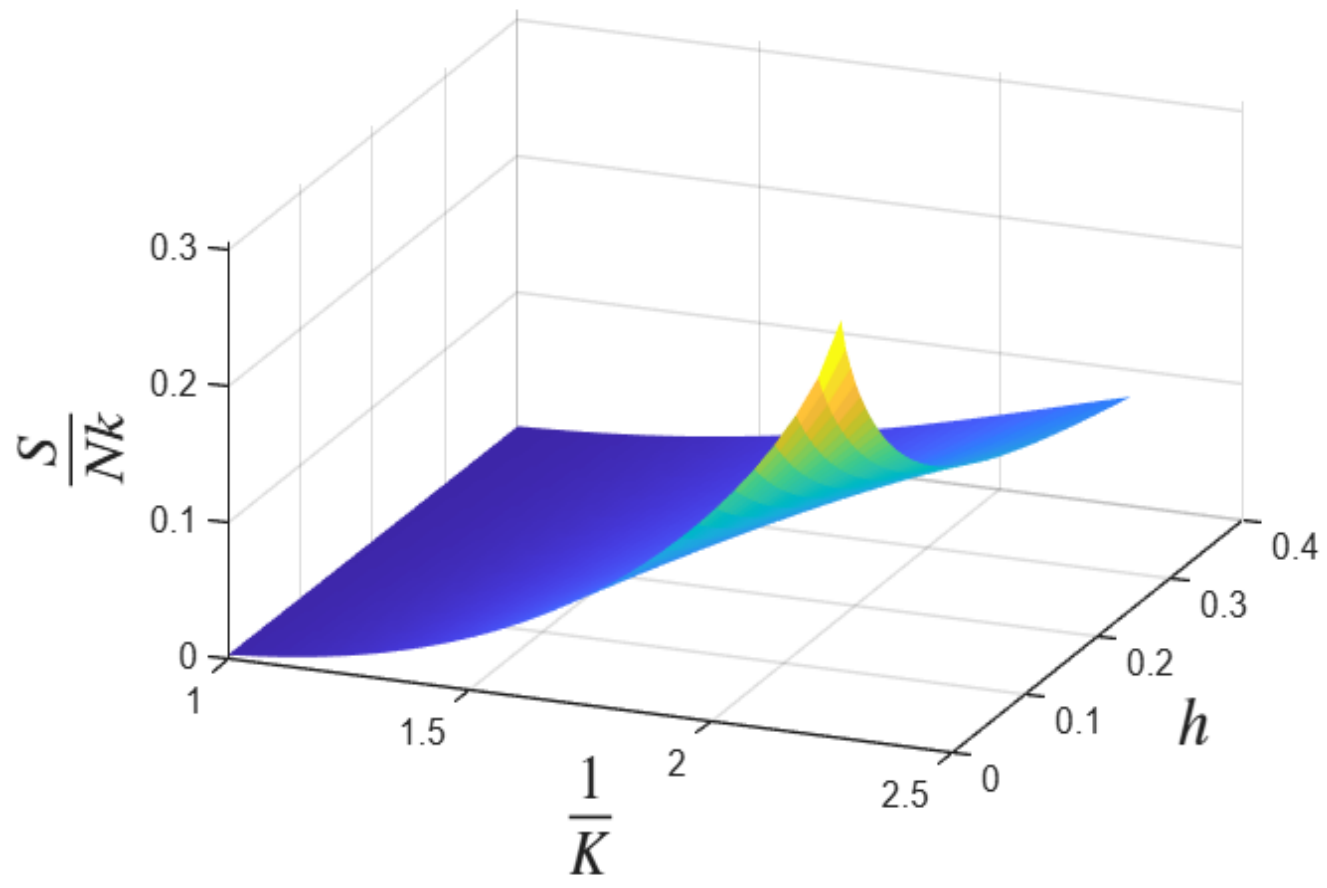

(a)

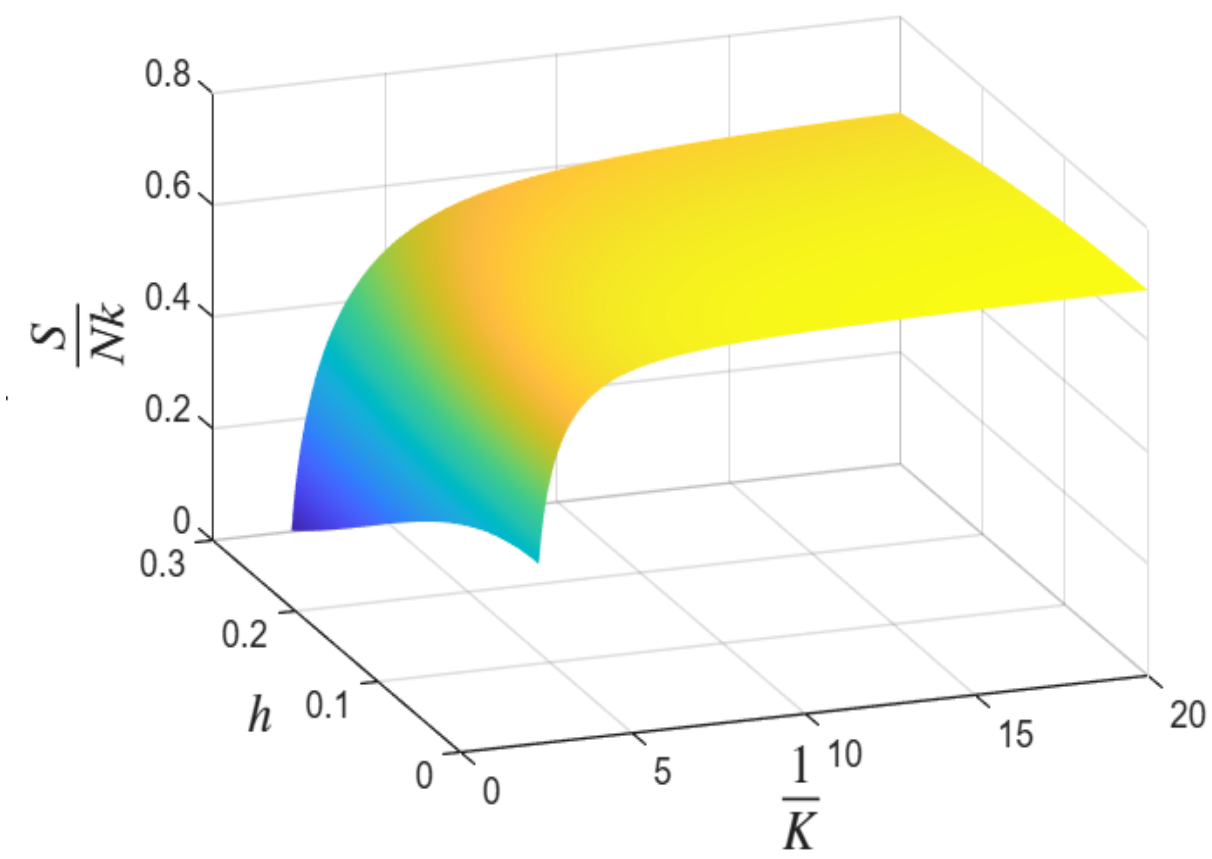

(b)

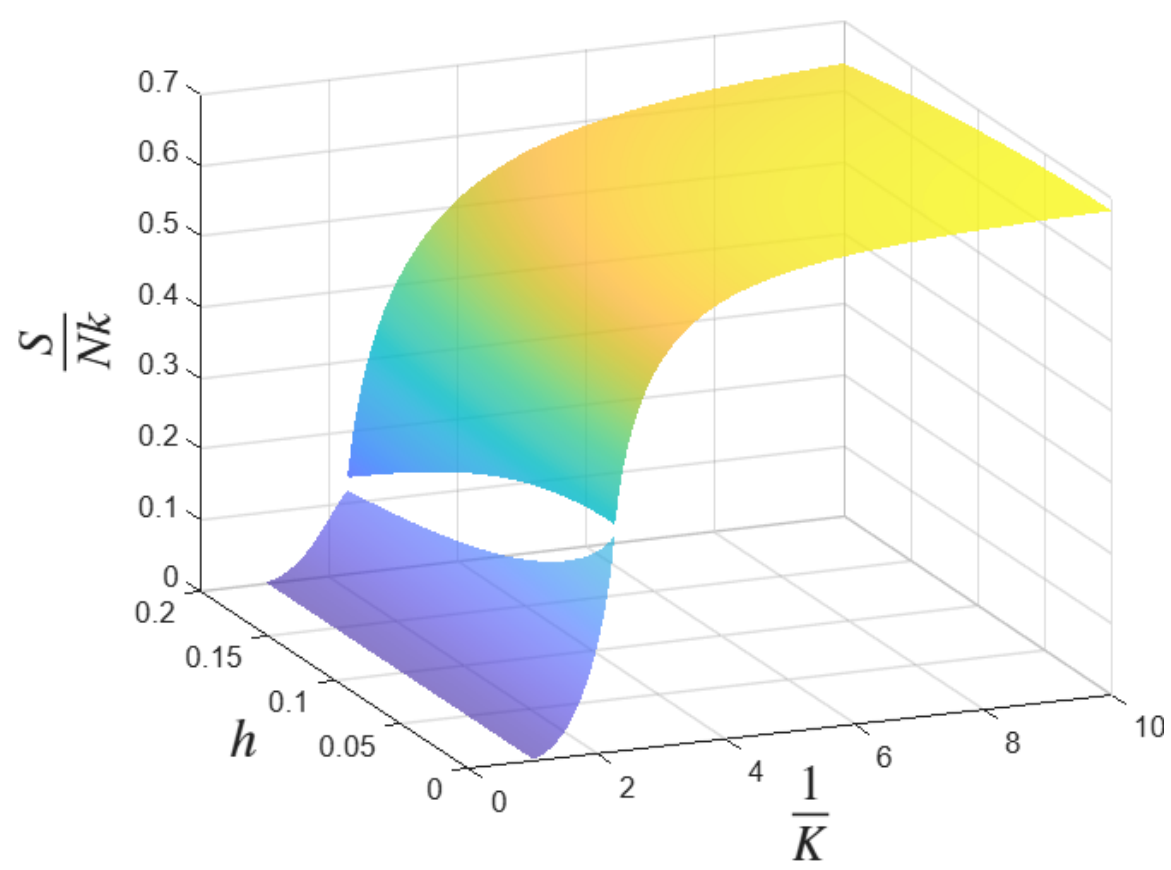

(c)

Figure 6: Dependence of entropy on the magnetic field and interaction energies:(a) $\frac{T}{T_c} \leq 1$ and (b) $\frac{T}{T_c} > 1$ and (c) both ranges of $\frac{T}{T_c}$

**Magnetization**

The formulation of spontaneous magnetization of two-dimensional Ising models has been a fascinating field of research in condensed matter physics. It is customary to define the spontaneous magnetization as [31]

$$M_0(T) = \lim_{h\to 0+} M(\mathrm{h})$$

wherein the thermodynamic limit and sequence of applying limits plays a crucial role. While equation (5) for $M\ (h = 0)$ was originally propounded by Onsager [3], the exact derivation was first provided by Yang [27]. In contrast to earlier approaches, we have deduced the magnetization directly from the partition function and it turns out to be an odd function of $h$ for both low and high temperatures. The low temperature series expansion for $M\ (h, K)$ using equation (26) yields all the coefficients up to $u^{12}$ [ 34] and that for high temperature pertaining to the zero-field susceptibility yields the coefficients till $tanh^4 K,$ [34] although additional terms can be derived at the cost of enhanced complexity (Table 1).

While the spontaneous magnetization expression at $h = 0$ is correctly obtained, the 'structure' of $\kappa\ (h = 0)$ is kept intact even when $h \neq 0$ viz

$$M = \left[1 - \frac{f(u,h)_{h\geq 0}}{\sinh^4(2K+h)}\right]^{\frac{1}{8}}$$

and

$$\kappa = \frac{2(1-M^8)^{\frac{1}{4}}}{1+(1-M^8)^{\frac{1}{2}}}$$

Figure 7a is in qualitative agreement with Fig 1 of Plischke and Mattis [35], obtained using finite-size calculations as well as Figure 6 of Stump [36], obtained using a microcanonical ensemble. Figure 7b depicts the variation of $M$ with $h$ at the critical temperature, along with the predicted variation of $M$ with $h^{\frac{1}{15}}$. The lines nearly coincide thus indicating the validity of equation (26) for the critical exponent $\delta$. For clarity, we have depicted the dependence of M on h and K for low and high temperatures in Fig 8a and Fig 8b respectively.

.

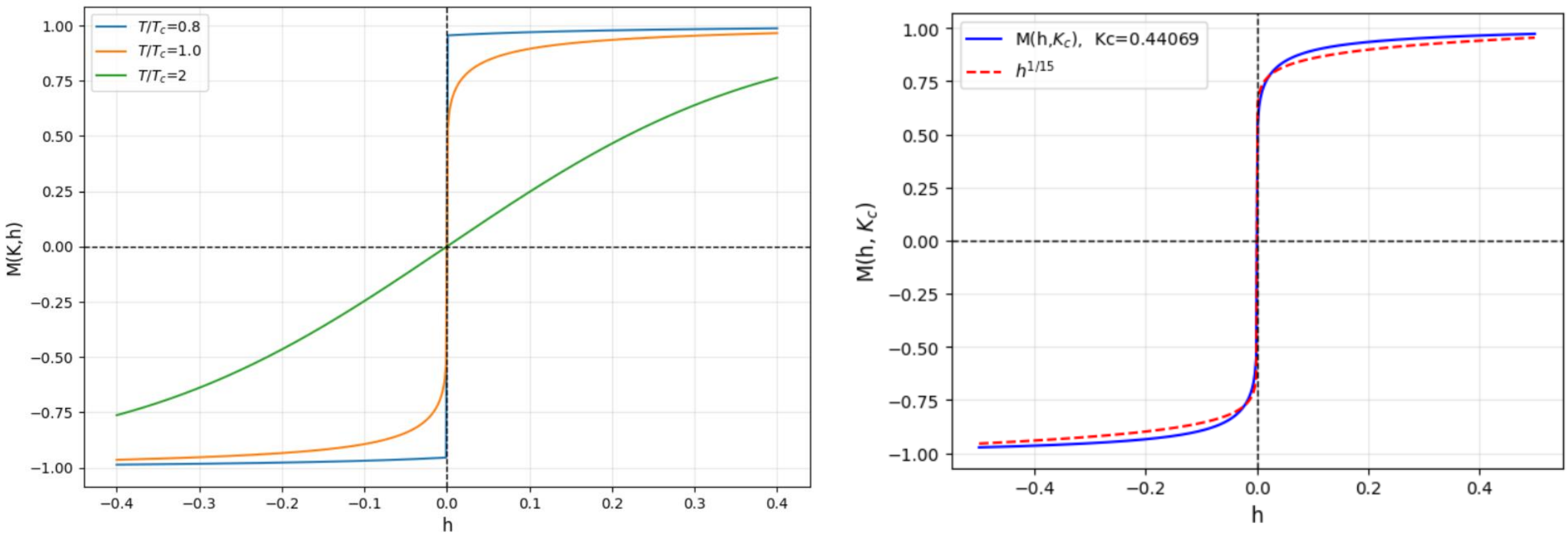

(a) (b)

Figure 7: Dependence of the magnetization on the magnetic field for (a) $\frac{T}{T_c}$ = 0.8, 1.0 and 2.0 and (b) $\frac{T}{T_c} = 1$.

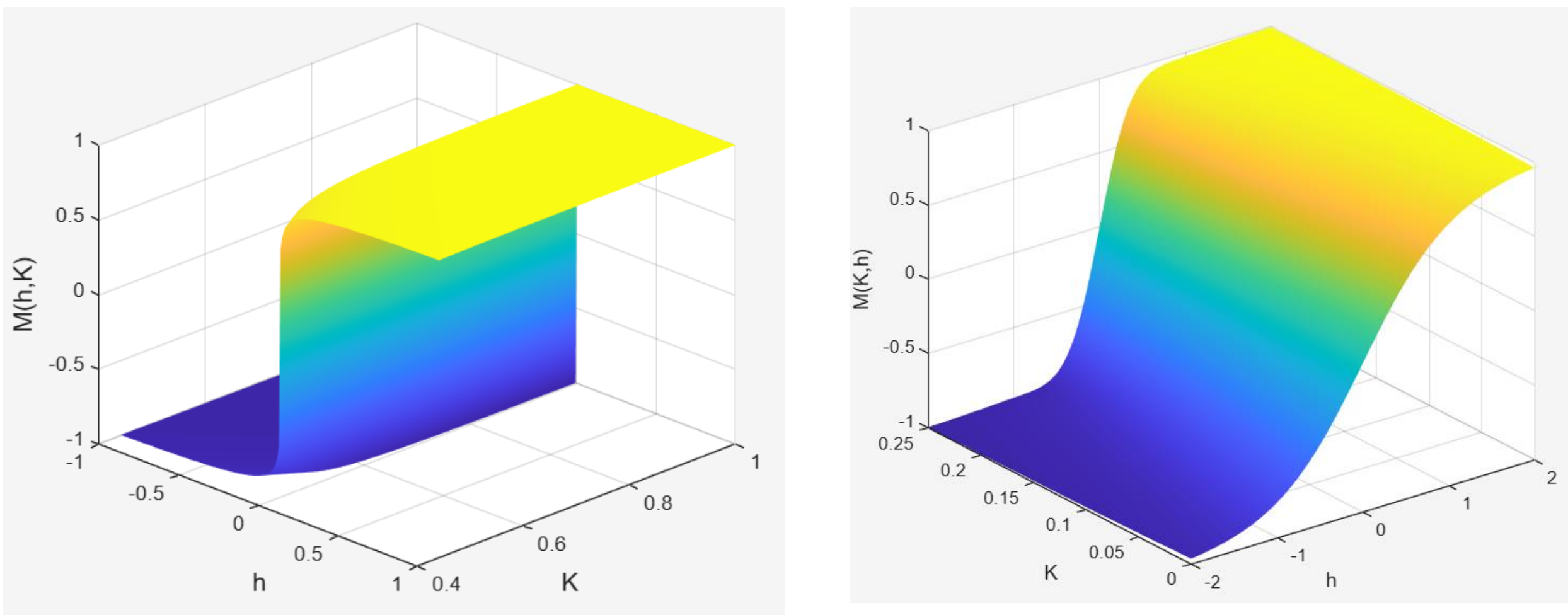

(a) (b)

Figure 8: Dependence of magnetization on the magnetic field and interaction energies

(a) $\frac{T}{T_c} \leq 1$ low temperature and (b) $\frac{T}{T_c} > 1$

## Susceptibility

The variation of the susceptibility with magnetic field and interaction energies is shown in Figure 9a (low temperatures) and Figure 9b (for high temperatures). In the latter case, the first entry in the third row of Table 1 was employed for the function for $g_1(h, K)$ which has yielded the zero-field susceptibility series till $tanh^4 K$ term. In Fig 9c, we have compared the zero-field susceptibility estimated here with the variation predicted by $M_0^{-14}$ *vis a vis* $\left(1 - \frac{T}{T_c}\right)^{-\frac{7}{4}}$ and the agreement is satisfactory.

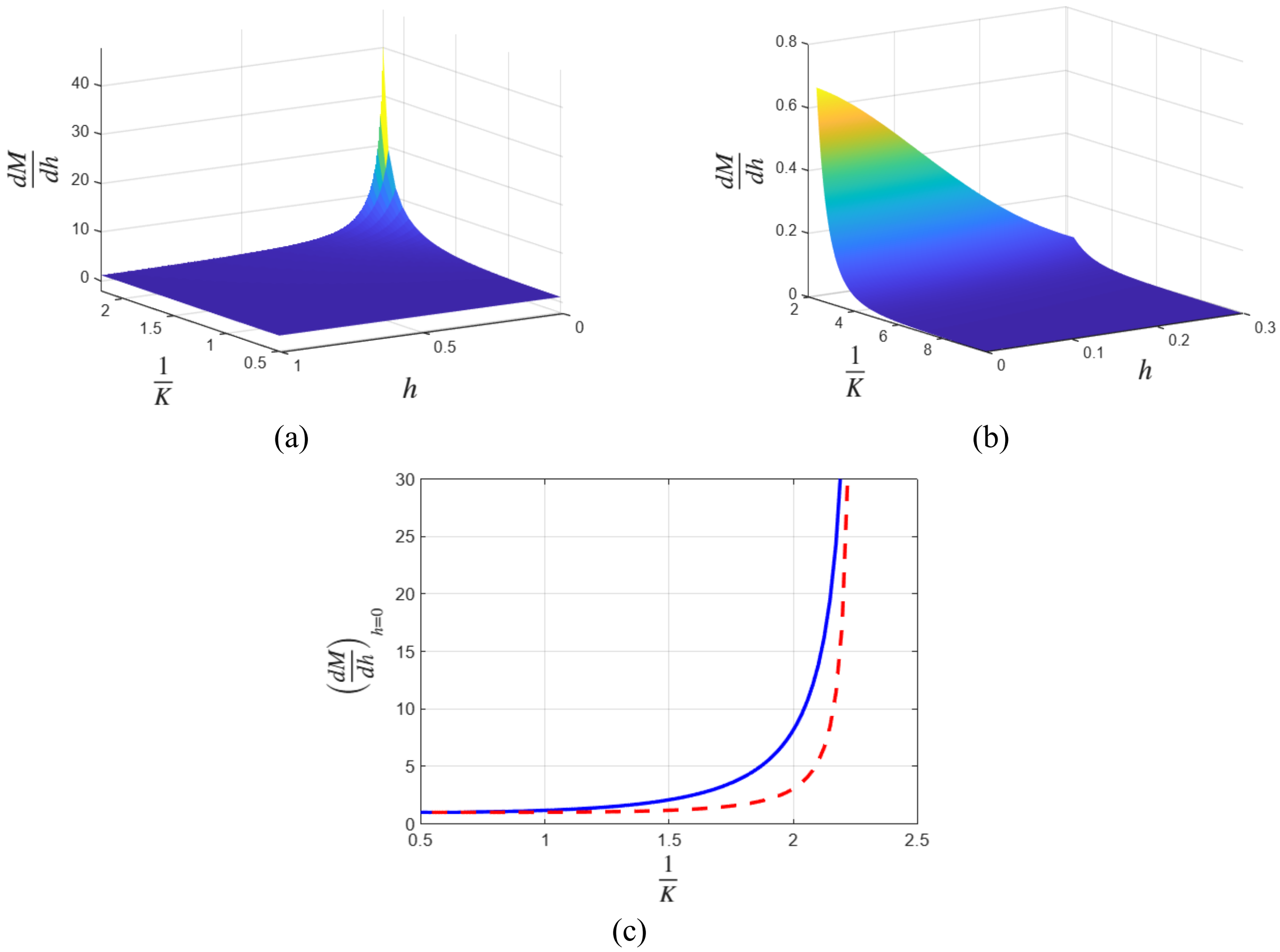

Figure 9: The variation of the susceptibility with the magnetic field and interaction energies: (a) $\frac{T}{T_c} \leq 1$ and (b) $\frac{T}{T_c} > 1$ and (c) dependence of the zero-field susceptibility on $\frac{1}{K}$ . The blue line denotes the values obtained by differentiation of equation (26) with respect to the magnetic field while the red line indicates the variation predicted by $M_0^{-14}$ *vis a vis* $\left(1 - \frac{T}{T_c}\right)^{-\frac{7}{4}}$.

As mentioned by McCoy and Wu[37] , the vanishing of the spontaneous magnetization as well as the maximum of the zero-field susceptibility occur at the same temperature viz $T_c$ . The origin of this behavior may be comprehended from an analogous dependence in the potential distribution at metal/electrolyte interfaces. The dipole potential ( $g_{dip}$) which is related to the order parameter of solvent molecules is an odd function of the charge density ($\sigma^M$) on the metal surface. The corresponding differential capacitance d ($g_{dip}$)/d$\sigma^M$ as well as its temperature coefficient is an even function of $\sigma^M$ . The maxima pertaining to the differential capacitance as well as the vanishing of the dipole potential both occur at the same charge density, consistent with the similar behaviour in two-dimensional Ising models as a function of temperature [38]. In fact, phase transitions at metal/electrolyte interfaces can be modelled using a generalized Ising model formalism [39].

There are two intriguing aspects of our methodology viz (i) two separate equations for the partition function, depending upon the temperature range and (ii) incorporation of empirical functions for deducing the non-zero magnetic field results. The former limitation can be surmounted by employing a two-point Pade' Approximant [ 40] to derive a single function valid for all temperatures. Since the magnetization exhibits different behaviour depending upon the ratio $\frac{T}{T_c}$ [28,31] and the amplitudes of zero-field susceptibility are dissimilar [41], two separate equations for the field-dependent partition functions may indeed be essential. The other limitation of employing 'ad hoc' functions at this *preliminary* methodology is warranted for obtaining the coefficients of the series expansions pertaining to the magnetization and susceptibility, for all temperature ranges. A comprehensive methodology to overcome these limitations is now under investigation.

**Summary**

By incorporating the external magnetic field within the graph-theoretical frame work of two-dimensional Ising models, the partition function, magnetization, internal energy, specific heat, susceptibility and entropy have been analytically obtained *albeit* with arbitrary polynomials for low and high temperatures. The similarity with Onsager' s exact solutions is maintained and truncated series expansions of various parameters are correctly reproduced.

Acknowledgements

The financial support work by the Mathematical Research Impact Centric Scheme (MATRICS) of SERB, Government of India is gratefully acknowledged. We thank the P.G. Senapathy Centre for Computing Resources, Indian Institute of Technology-Madras Chennai.

**Appendix A**

In this Appendix, the eigenvalues of the matrix T are provided.

$$\lambda_1 = \beta(K,h)\left[\frac{1}{2}\left(cos(q_2)+ cos(q_1)\right)-\frac{1}{2}\sqrt{\left(cos(q_2)+ cos(q_1)\right)^2 - 4} - \frac{1}{2}\sqrt{2\left(cos(q_2)+ cos(q_1)\right)^2 - \frac{8\left(cos(q_2)+ cos(q_1)\right)^3 - 32\left(cos(q_2)+ cos(q_1)\right)}{4\left(\left(cos(q_2)+ cos(q_1)\right)^2 - 4\right)^{\frac{1}{2}}}}\right]$$

$$\lambda_2 = \beta(K,h)\left[\frac{1}{2}\left(cos(q_2)+ cos(q_1)\right)-\frac{1}{2}\sqrt{\left(cos(q_2)+ cos(q_1)\right)^2 - 4} + \frac{1}{2}\sqrt{2\left(cos(q_2)+ cos(q_1)\right)^2 - \frac{8\left(cos(q_2)+ cos(q_1)\right)^3 - 32\left(cos(q_2)+ cos(q_1)\right)}{4\left(\left(cos(q_2)+ cos(q_1)\right)^2 - 4\right)^{\frac{1}{2}}}}\right]$$

$$\lambda_3 = \beta(K,h)\left[\frac{1}{2}\left(cos(q_2) + cos(q_1)\right) + \frac{1}{2}\sqrt{\left(cos(q_2) + cos(q_1)\right)^2 - 4}\right.$$

$$\left. - \frac{1}{2}\sqrt{2\left(cos(q_2) + cos(q_1)\right)^2 - \frac{8\left(cos(q_2) + cos(q_1)\right)^3 - 32\left(cos(q_2) + cos(q_1)\right)}{4\left(\left(cos(q_2) + cos(q_1)\right)^2 - 4\right)^{\frac{1}{2}}}}\right]$$

$$\lambda_4 = \beta(K,h)\left[\frac{1}{2}\left(cos(q_2) + cos(q_1)\right) + \frac{1}{2}\sqrt{\left(cos(q_2) + cos(q_1)\right)^2 - 4}\right.$$

$$\left. + \frac{1}{2}\sqrt{2\left(cos(q_2) + cos(q_1)\right)^2 - \frac{8\left(cos(q_2) + cos(q_1)\right)^3 - 32\left(cos(q_2) + cos(q_1)\right)}{4\left(\left(cos(q_2) + cos(q_1)\right)^2 - 4\right)^{\frac{1}{2}}}}\right]$$

## Appendix B

In this Appendix, we convert the double integral (17) into a single integral using standard techniques and deduce the partition function in a simpler form, in order to demonstrate the similarity with Onsager's exact solution.

$$I = \frac{1}{8\pi^2}\int_0^{2\pi}\int_0^{2\pi} \ln\left\{1 - \frac{2\beta(K,h)(1-\beta(K,h)^2)}{(1+\beta(K,h)^2)^2}\ (cosq_1 + cosq_2)\right\} dq_1\, dq_2$$

Upon re-defining the variables as

$$u = \frac{q_1 + q_2}{2}, v = \frac{q_1 - q_2}{2}$$

$$I = \frac{1}{4\pi^2}\int_0^{\pi}\int_0^{2\pi} \ln\left\{1 - \frac{4\beta(K,h)(1-\beta(K,h)^2)}{(1+\beta(K,h)^2)^2}\cos(u)\cos(v)\right\} dvdu$$

$$I = \frac{1}{2\pi}\int_0^{\pi} \ln\left[\frac{1 + \sqrt{1 - \left(\frac{4\beta(K,h)(1-\beta(K,h)^2)}{(1+\beta(K,h)^2)^2}\right)^2 \cos^2(u)}}{2}\right] du$$

Using another variable change as $u = \frac{\pi}{2} - \varphi$

$$I = \frac{1}{\pi}\int_0^{\frac{\pi}{2}} \ln\left[\frac{1 + \sqrt{1 - \left(\frac{4\beta(K,h)(1-\beta(K,h)^2)}{(1+\beta(K,h)^2)^2}\right)^2 \sin^2\varphi}}{2}\right] d\varphi$$

Hence the partition function becomes

$$\frac{1}{N}\ lnQ_{(h\geq 0)} = \frac{1}{2}\ln\left[\frac{2\alpha(K,h)(1+\beta(K,h)^2)^2}{(1-x^2)^2(1-y^2)^{\frac{1}{2}}}\right]$$
$$+\frac{1}{\pi}\int_0^{\frac{\pi}{2}} ln\left[1+\sqrt{1-\left(\frac{4\beta(K,h)(1-\beta(K,h)^2)}{(1+\beta(K,h)^2)^2}\right)^2 sin^2\varphi}\right]d\varphi$$

which is equation (18) of the text.

## Appendix C

In this Appendix, we report the internal energy ($U$), specific heat ($C_v$) and entropy (S) as a function of $h$ and $K$ . These equations reduce to the corresponding equations when $h = 0$. For brevity, we rewrite $f(u,h)_{h\geq 0}$ as $f(T)$, $g(u,h)_{h\geq 0}$ as $g(T)$ and $\kappa(u,h)_{h\geq 0}$ as $\kappa(T)$.

### Internal energy

The internal energy when $h\ \geq 0$ is deduced from the partition function as

$$\frac{U}{NJ} = \frac{kT^2}{J}\left\{\frac{g'(T)}{g(T)} + \frac{k\,T^2 f'(T) - 2\left(f(T)\right)^{\frac{1}{2}}(H\ +\ 2\,J)sinh\left(\frac{2\,(H\ +\ 2\,J)}{k\,T}\right)}{4\,k\,T^2\left(f(T)\right)^{\frac{1}{2}}\left(\left(f(T)\right)^{\left(\frac{1}{2}\right)} +\ sinh^2\left(\frac{H\ +\ 2\,J}{k\,T}\right)\right)} + \frac{1}{\pi}\left\{\frac{\kappa'(T)\left(\frac{\pi}{2} -\ K(\kappa(T)^2)\right)}{\kappa(T)}\right\}\right\}$$

$K(\kappa(T)^2)$ denotes the complete elliptic integral of the first kind

Here, $g(T)_{h=0} = 1, g'(T)_{h=0} = 0, g''(T)_{h=0}, f(T)_{h=0} = 1, f'(T)_{h=0} = 0, f''(T)_{h=0} = 0$

At H =0, the above equation leads to

$$\frac{U}{NJ} = \frac{kT^2}{J}\left\{-\frac{2\,J\ tanh\left(\frac{2\,J}{k\,T}\right)}{k\,T^2} + \frac{1}{\pi}\left\{\frac{\kappa'(T)\left(\frac{\pi}{2} -\ K(\kappa(T)^2)\right)}{\kappa(T)}\right\}\right\}$$

### Specific heat

The specific heat when h ≥ 0 follows from the equation

$$\frac{C_V}{Nk} = \frac{\partial}{\partial T}\left(T^2\frac{d(lnQ)}{dT}\right)$$

$$\frac{\partial}{\partial T}\left(T^2\frac{d(lnQ)}{dT}\right)$$

$$=\frac{k\,T\,\left(T\,g(T)g''(T)-\,T\,g'(T)^2+\,2\,g(T)g'(T)\right)}{g(T)^2}$$

$$+\frac{\left[\begin{array}{c} 2\,k^2T^4f(T)f''(T)cosh\left(\frac{2\,H}{k\,T}+\frac{4\,J}{k\,T}\right)+\,4\,k^2T^4f(T)^{\frac{3}{2}}f''(T)-\,2\,k^2T^4f(T)f''(T) \\ -\,k^2T^4f'(T)^2cosh\left(\frac{2\,H}{k\,T}+\frac{4\,J}{k\,T}\right)+\,8\,H\,k\,T^2f(T)f'(T)sinh\left(\frac{2\,H}{k\,T}+\frac{4\,J}{k\,T}\right)+\,16\,J\,k\,T^2f(T)f'(T)sinh\left(\frac{2\,H}{k\,T}+\frac{4\,J}{k\,T}\right) \\ +\,k^2T^4f'(T)^2-\,4\,k^2T^4\sqrt{f(T)}f'(T)^2+\,16\,H^2f(T)^2cosh\left(\frac{2\,H}{k\,T}+\frac{4\,J}{k\,T}\right)-\,8\,H^2f(T)^{\frac{3}{2}}cosh\left(\frac{2\,H}{k\,T}+\frac{4\,J}{k\,T}\right) \\ +\,8\,H^2f(T)^{\frac{3}{2}}+\,64\,J^2f(T)^2cosh\left(\frac{2\,H}{k\,T}+\frac{4\,J}{k\,T}\right)-\,32\,J^2f(T)^{\frac{3}{2}}cosh\left(\frac{2\,H}{k\,T}+\frac{4\,J}{k\,T}\right)+\,16\,H\,k\,T\,f(T)^2sinh\left(\frac{2\,H}{k\,T}+\frac{4\,J}{k\,T}\right) \\ +\,32\,J\,k\,T\,f(T)^2sinH\left(\frac{2\,H}{k\,T}+\frac{4\,J}{k\,T}\right)-\,8\,H\,k\,T\,f(T)^{\frac{3}{2}}sinH\left(\frac{2\,H}{k\,T}+\frac{4\,J}{k\,T}\right)-\,16\,J\,k\,T\,f(T)^{\frac{3}{2}}sinh\left(\frac{2\,H}{k\,T}+\frac{4\,J}{k\,T}\right) \\ +\,4\,H\,k\,T\,f(T)^{\frac{3}{2}}sinh\left(\frac{4\,H}{k\,T}+\frac{8\,J}{k\,T}\right)+\,8\,J\,k\,T\,f(T)^{\frac{3}{2}}sinh\left(\frac{4\,H}{k\,T}+\frac{8\,J}{k\,T}\right)+\,64\,H\,J\,f(T)^2cosh\left(\frac{2\,H}{k\,T}+\frac{4\,J}{k\,T}\right) \\ -\,32\,H\,J\,f(T)^{\frac{3}{2}}cosh\left(\frac{2\,H}{k\,T}+\frac{4\,J}{k\,T}\right)+\,32\,H\,J\,f(T)^{\frac{3}{2}}+\,32\,J^2f(T)^{\frac{3}{2}} \end{array}\right]}{4\,k^2T^4f(T)^{\frac{3}{2}}\left(2\,\sqrt{f(T)}+\,cosh\left(\frac{2\,H}{k\,T}+\frac{4\,J}{k\,T}\right)-\,1\right)^2}$$

$$+\frac{\left[T\left(\begin{array}{c}\frac{1}{2}T\,\left(\pi\,-\,2\,K\left(k(T)^2\right)\right)k''(T)+\left(\pi\,-\,2\,K\left(k(T)^2\right)\right)k'(T) \\ +\frac{T\,k'(T)^2\left(k(T)^2K\left(k(T)^2\right)-\,K\left(k(T)^2\right)+\,E\left(k(T)^2\right)\right)}{k(T)\left(k(T)^2-\,1\right)}\end{array}\right)\right]}{\pi}$$

Here $K(\kappa(T)^2)$ is the complete elliptic integral of the first kind and $E(k(T)^2)$ is the complete elliptic integral of the second kind.

When H =0, the above equation leads to

$$\frac{C_V}{Nk}=-\frac{T^2\kappa(T)K(\kappa(T)^2)\kappa''(T)}{\pi\,(\kappa(T)^2-\,1)}+\frac{T^2K(\kappa(T)^2)\kappa''(T)}{\pi\,\kappa(T)(\kappa(T)^2-\,1)}+\frac{2\,T^2K(\kappa(T)^2)\kappa'(T)^2}{\pi\,(\kappa(T)^2-\,1)}-\frac{2\,T^2K(\kappa(T)^2)\kappa'(T)^2}{\pi\,\kappa(T)^2(\kappa(T)^2-\,1)}$$
$$-\frac{2\,T\,\kappa(T)K(\kappa(T)^2)\kappa'(T)}{\pi\,(\kappa(T)^2-\,1)}+\frac{2\,T\,K(\kappa(T)^2)\kappa'(T)}{\pi\,\kappa(T)(\kappa(T)^2-\,1)}-\frac{4\,J^2sech^2\left(\frac{2\,J}{k\,T}\right)}{k^2T^2(\kappa(T)^2-\,1)}+\frac{4\,J^2\kappa(T)^2sech^2\left(\frac{2\,J}{k\,T}\right)}{k^2T^2(\kappa(T)^2-\,1)}$$
$$+\frac{T^2\kappa(T)\kappa''(T)}{2\,(\kappa(T)^2-\,1)}-\frac{T^2\kappa''(T)}{2\,\kappa(T)(\kappa(T)^2-\,1)}+\frac{T^2\kappa'(T)^2}{2\,\kappa(T)^2(\kappa(T)^2-\,1)}-\frac{T^2\kappa'(T)^2}{2\,(\kappa(T)^2-\,1)}$$
$$+\frac{T^2\kappa'(T)^2E(\kappa(T)^2)}{\pi\,\kappa(T)^2(\kappa(T)^2-\,1)}+\frac{T\,\kappa(T)\kappa'(T)}{\kappa(T)^2-\,1}-\frac{T\,\kappa'(T)}{\kappa(T)(\kappa(T)^2-\,1)}$$

which is identical with the result from the exact solution.

**Entropy**

The equation for entropy when H ≥ 0 is given by

$$\frac{S}{Nk}=\ln(g(T))+\ln\left\{2^{\frac{1}{2}}\left(f(T)^{\frac{1}{2}}+sinh^2\left(\frac{H}{kT}+\frac{2J}{kT}\right)\right)^{\frac{1}{2}}\right\}+\frac{1}{\pi}\int_0^{\frac{\pi}{2}}\ln\left(1+\sqrt{1-\kappa(T)^2\sin^2\varphi}\right)d\varphi$$

$$+T\left\{\frac{g'(T)}{g(T)}+\frac{k\,T^2f'(T)-\,2\,(f(T))^{\frac{1}{2}}(H\,+\,2\,J)sinh\left(\frac{2\,(H\,+\,2\,J)}{k\,T}\right)}{4\,k\,T^2(f(T))^{\frac{1}{2}}\left((f(T))^{\left(\frac{1}{2}\right)}+\,sinh^2\left(\frac{H\,+\,2\,J}{k\,T}\right)\right)}+\frac{1}{\pi}\left\{\frac{\kappa'(T)\left(\frac{\pi}{2}-\,K(\kappa(T)^2)\right)}{\kappa(T)}\right\}\right\}$$

When $H = 0$,the above equation reduces to

$$\frac{S}{k} = ln\left[2^{\frac{1}{2}} cosh\left(\frac{2J}{kT}\right)\right] + \frac{1}{\pi}\int_0^{\frac{\pi}{2}} ln\left(1 + \sqrt{(1 - \kappa^2 \sin^2 \phi)}\right) d\phi$$
$$+ T\left\{-\frac{2J\, tanh\left(\frac{2J}{kT}\right)}{k\,T^2} + \frac{1}{\pi}\left\{\frac{\kappa'(T)\left(\frac{\pi}{2} - K(\kappa(T)^2)\right)}{\kappa(T)}\right\}\right\}$$

which is identical with the equation arising from Onsager's exact solution.